\documentclass[aps,prd,twocolumn,floats,floatfix,nofootinbib,longbibliography,10pt]{revtex4-2}

\usepackage{graphicx}
\usepackage{amsmath,amsfonts}
\usepackage{dcolumn}
\usepackage{hyperref}

\mathchardef\minus = "002D

\newcommand{\swS}[5][]{{}_{{}_{#2}}S^{#1}_{#3}(#4;#5)}  
\newcommand{\scA}[4][]{{}_{{}_{#2}}A^{#1}_{#3}(#4)} 
\newcommand{\scAnorm}[4][]{{}_{{}_{#2}}\bar{A}^{#1}_{#3}(#4)} 

\newcolumntype{.}{D{.}{.}{-1}}
\newcolumntype{d}[1]{D{.}{.}{#1}}

\begin{document}

\title{New total transmission modes of the {K}err geometry with {S}chwarzschild limit frequencies at complex infinity}

\author{Gregory B. Cook}\email{cookgb@wfu.edu}
\affiliation{Department of Physics, Wake Forest University,
		 Winston-Salem, North Carolina 27109}
\author{Suhan Lu}\email{lus19@wfu.edu}
\affiliation{Department of Physics, Wake Forest University,
     Winston-Salem, North Carolina 27109}

\date{\today}

\begin{abstract}
In addition to the well-known quasinormal modes, the gravitational modes of the Kerr geometry also include sets of total-transmission modes.  Each mode can be considered as an element of a sequence of modes parameterized by the angular momentum of the black hole.  One family of gravitational total-transmission modes of Kerr have been known for some time.  Modes in this family connect to a Schwarzschild limit where the mode frequency is finite and purely imaginary.  Recently, what was thought to be an additional branch of this original family of modes was discovered.  However, this new branch is actually a part of one of two entirely new families of total-transmission modes.  Modes in these new families, surprisingly, connect to a Schwarzschild limit where the mode frequencies exist at complex infinity.  We have numerically constructed full sets of sequences of gravitational total-transmission modes for harmonic indices $\ell=[2,8]$.  Using these numerical sequences, we have been able to construct analytic asymptotic expansions for the mode frequencies and their associated separation constants.  The asymptotic expansion for the separation constant used in constructing the total-transmission modes seems to be valid for general complex values of the oblateness parameter.
\end{abstract}

\maketitle

\section{Introduction}
\label{sec:introduction}

The total transmission modes (TTMs) of the Kerr geometry\cite{kerr-1963} represent linear perturbations of the Kerr metric that effectively travel radially through the spacetime without reflection.  There are two types of TTMs based on the behavior of modes at the black-hole horizon and at spatial infinity.  Left TTMs (TTM${}_{\rm L}$s) are modes which are forbidden from traveling into the black hole, and for which disturbances are not allowed to enter the system from infinity.  Essentially, TTM${}_{\rm L}$s represent disturbances that can move from the vicinity of the black hole toward spatial infinity without reflections that would propagate back into the black hole.  Right TTMs (TTM${}_{\rm R}$s) switch both boundary conditions.  Essentially, TTM${}_{\rm R}$s represent disturbances that can move from spatial infinity toward the black hole without reflections that would propagate back toward spatial infinity.  The TTMs, together with the quasinormal modes (QNMs), represent the full set of modes of the Kerr geometry.  The QNMs have been more thoroughly explored\cite{berti-QNM-2009,QNM-Review-2011} than the TTMs and are of more immediate physical relevance as they are used to model the ring-down portion of a gravitational waveform, for example from a binary-black-hole merger\cite{GW150914-2016,GW151226-2016,GW170104-2017,GW170814-2017}.

The TTMs were first explored, in the context of algebraically special modes, by Wald\cite{wald-1973} and in more detail by Chandrasekhar\cite{chandra-1984}.  In general, the mode frequencies, $\omega$, are functions of the angular momentum of the black hole, and must be determined numerically\cite{cook-zalutskiy-2014,cook-et-al-2018} or through analytic approximations\cite{andersson-1994,KeshetNeitzke2008}.  The modes and their frequencies are conveniently parameterized by the dimensionless angular momentum $\bar{a}=a/M$, where $M$ is the mass of the black hole and $J=aM$ is the black hole's angular momentum.  In the Schwarzschild limit, $\bar{a}=0$, the TTMs and their complex frequencies $\omega$ can be determined analytically.  The TTM frequencies $\omega(\bar{a})$ take on the purely imaginary values 
\begin{equation}\label{eqn:alg-spec-sch}
  M\omega_{\ell m0}(0) = -\frac{i}{12}(\ell-1)\ell(\ell+1)(\ell+2).
\end{equation}
Here, the index $\ell$ is a harmonic mode index, $m$ is the azimuthal mode index, and the third index(set to zero here) is an overtone index used to differentiate unique modes that have the same $\ell$ and $m$.  In general the mode frequencies $\omega(\bar{a})$ are complex and the TTM${}_{\rm L}$s and TTM${}_{\rm R}$s share the same set of mode frequencies, although the modes themselves differ.

The first numerical results for the mode frequencies of the TTMs were included in Chandrasekhar's investigation of the algebraically special perturbations of Kerr\cite{chandra-1984}.  There, a few values were given for each of the five modes for $\ell=2$.  Subsequently, these algebraically special perturbations were recognized as TTMs\cite{andersson-1994}.  More detailed plots for the $\ell=2$ modes were shown by Onozawa\cite{onozawa-1997}.  More recently, one of us reported the mode frequencies for both $\ell=2$ and $\ell=3$\cite{cook-zalutskiy-2014}, and then for $\ell=2$ through $\ell=7$\cite{cook-et-al-2018}.  In the latter work, a previously unknown branch of the $m=0$ TTMs was explored\cite{cook-et-al-2018}.  This unknown branch was discovered because of a subtle behavior that may occur when the roots of the indicial equation differ by an integer for the local Frobenius solution at the event horizon.  Certain QNM solutions with mode frequencies on the negative imaginary axis turn out to actually be simultaneously both a QNM and a TTM${}_{\rm L}$\cite{cook-zalutskiy-2016b}.  The existence of a discrete set of simultaneous QNM and TTM${}_{\rm L}$ solutions led to the discovery of a set of continuous sequences of TTM solutions with the unusual behavior that the Schwarzschild limit of these sequence did not have a finite mode frequency as in Eq.~(\ref{eqn:alg-spec-sch}), but instead the limiting mode frequency was at $M\omega=-i\infty$.

The set of TTMs that were originally studied and have a Schwarzschild limit with finite mode frequencies given by Eq.~(\ref{eqn:alg-spec-sch}) form an interesting family of solutions that will be distinguished by an overtone index of $n=0$.  As will be discussed in more detail below, it is known that for any QNM or TTM solution with mode frequency $\omega_{\ell mn}$, there also exists another solution with mode frequency $-\omega^*_{\ell(-m)n}$.  These two modes are often referred to as mirror modes, and in general the modes form two distinct sets of mode solutions which are frequently labeled as $\omega^+_{\ell mn}$ and $\omega^-_{\ell mn}=-\left(\omega^+_{\ell(-m)n}\right)^*$.  The first known family of TTMs has the interesting behavior that its mirror-mode solutions are degenerate.  That is, $\omega^-_{\ell m0}=\omega^+_{\ell m0}$.  Because of this, a common perspective is to assume that the $\omega^+_{\ell m0}$ TTMs only exist for $m\ge0$ and the $\omega^-_{\ell m0}$ TTMs only exist for $m\le0$, but this restriction is entirely a matter of bookkeeping.

Also interesting for this first family of TTMs is that all of the $m=0$ mode sequences exhibit a single point along each sequence where $\omega^\pm_{\ell00}$ becomes $C^0$.  That is, at some value of $\bar{a}\equiv\hat{a}_\ell$, the sequence of frequencies is continuous, but not smooth.  For $0\le\bar{a}\le\hat{a}_\ell$, the sequences of mode frequencies are smooth and purely imaginary.  This portion of each sequence is denoted in overtone-multiplet notation as $n=0_0$.  For $\hat{a}_\ell<\bar{a}\le1$, the sequences of mode frequencies are again smooth, but are now complex valued.  This portion of each sequence is denoted in overtone-multiplet notation as $n=0_1$.  The previously unknown branch of the $m=0$ TTMs for each $\ell$, explored in Ref.~\cite{cook-et-al-2018}, also has purely imaginary frequencies and exists for $0\le\bar{a}\le\hat{a}_\ell$ as does each of the $n=0_0$ sequences.  However, each new branch has a Schwarzschild limit at $M\omega_{\ell00_2}=-i\infty$, and $|\omega_{\ell00_0}(\bar{a})|\le|\omega_{\ell00_2}(\bar{a})|$ with equality only at $\bar{a}=\hat{a}_\ell$.

The previously unknown branch of the $m=0$ TTMs in the preceding paragraph was labeled in Ref.~\cite{cook-et-al-2018} as $n=0_2$, a third member of the $m=0$, $n=0$ overtone multiplet.  This was a reasonable choice when this new sequence of solutions was the only addition to the set of known TTMs.  However, we have now found numerous previously unknown sequences of TTMs spanning all allowed values of $\ell$ and $m$.  These new sequences were found by brute-force methods described in the main text, and full sequences of numerical solutions were constructed for values of $2\le\ell\le8$.  Common to all of these new sequences is that they have a Schwarzschild limit with mode frequency somewhere along complex infinity.  Examining the behavior of these new sequences, it is clear that they divide into two new families of TTMs which will be denoted by $n=1$ and $n=2$ overtone values.  One of these new families, which we label by $n=1$, shares the behavior of the original family that its mirror-mode solutions are degenerate.  But, the other family, labeled by $n=2$, displays a full complement of modes and mirror modes.  Finally, it is clear that the sequences which had been labeled by the $n=0_2$ overtone multiplet should actually belong to the new $n=1$ family of TTMs.  So, some relabeling of solutions will be necessary.

The proper labeling of these new sequences is not a trivial exercise.  The harmonic index $\ell$ labeling all modes is somewhat arbitrary.  For general modes that have a spherical harmonic decomposition, the meaning of the index $\ell$ is well motivated because the eigenvalues of the spherical harmonics are known analytically in terms of $\ell$, and $L\equiv\ell-|m|$ specifies the number of zero crossings of the associated Legendre polynomials $P_{\ell m}(\cos\theta)$ expressing the $\theta$ dependence of the mode.  For general modes that can be decomposed in terms of spin-weighted spherical harmonics, the situation is similarly well defined.  For TTMs (and QNMs) that have a Schwarzschild limit where the mode frequency is finite and $|a\omega|=0$, a choice of $\ell$ for the entire sequence can be made at this limit where the angular decomposition is realized by spin-weighted spherical harmonics.  When $|a\omega|>0$, the harmonic decomposition of modes of the Kerr geometry is made in terms of the spin-weighted {\em spheroidal} harmonics\cite{teukolsky-1973} and associating a specific value of $\ell$ with a specific eigensolution is no longer so clearly defined\cite{VickersCook2022}.

In order to uniquely fix the labeling of the new TTM sequences, we have constructed from our numerical solutions an analytic asymptotic expansion for the mode frequencies and for the eigenvalues of the spin-weighted spheroidal harmonics.  This expansion is again in the Schwarzschild limit of the sequences, but for the new TTMs the Schwarzschild limit exists where $|a\omega|\to\infty$.  These asymptotic expansions provide a unique, well-motivated labeling of modes for both the harmonic index $\ell$ and the overtone family $n$.\footnote{There is no ambiguity related to the azimuthal index $m$.}

This paper proceeds as follows.  In Sec.~\ref{sec:methods}, we will briefly review the methods used to construct TTMs of the Kerr geometry, including our approach for constructing asymptotic expansions from the numerical sequences of solutions.  In Sec.~\ref{sec:numerical_results}, we will explore the general behavior of the mode frequencies of the numerical TTM solutions, and we will derive the analytic asymptotic expansions for the mode frequencies of the two new families of TTMs along with an analytic expansion for the separation constant of the angular Teukolsky equation.  Finally, in Sec.~\ref{sec:discussion} we will discuss the results.

\section{Methods}
\label{sec:methods}

Total-transmission modes of the Kerr geometry can be obtained by solving the Teukolsky master equation with appropriate boundary conditions.  Our approach for obtaining the TTMs of the Kerr geometry is outlined in detail in Refs.~\cite{cook-et-al-2018,cook-zalutskiy-2014,cook-zalutskiy-2016b}.  Here, we will outline only the most important aspects of the approach.

In vacuum, the Teukolsky master equation separates using
\begin{equation}\label{eqn:Teukolsky_separation_form}
  {}_s\psi(t,r,\theta,\phi) = e^{-i\omega{t}} e^{im\phi}S(\theta)R(r).
\end{equation}
The radial function $R(r)$ then satisfies the radial Teukolsky equation
\begin{subequations}
\begin{align}\label{eqn:radialR:Diff_Eqn}
\Delta^{-s}\frac{d}{dr}&\left[\Delta^{s+1}\frac{dR(r)}{dr}\right]
 \\
&+ \left[\frac{K^2 -2is(r-M)K}{\Delta} + 4is\omega{r} - \lambdabar\right]R(r)=0,
\nonumber
\end{align}
where
\begin{align}
  \Delta &\equiv r^2-2Mr+a^2, \\
  K &\equiv (r^2+a^2)\omega - am, \\
\label{eqn:lambdabar def}
  \lambdabar &\equiv \scA{s}{\ell{m}}{a\omega} + a^2\omega^2 - 2am\omega,
\end{align}
\end{subequations}
and Boyer-Lindquist coordinates are used.  $\scA{s}{\ell{m}}{a\omega}$ is the angular separation constant associated with the angular Teukolsky equation governing $S(\theta)$.  With $x=\cos\theta$, the function $S(\theta)=\swS{s}{\ell{m}}{x}{a\omega}$ is the spin-weighted spheroidal function satisfying
\begin{align}\label{eqn:swSF_DiffEqn}
\partial_x \Big[ (1-x^2)\partial_x [\swS{s}{\ell{m}}{x}{c}]\Big] & \nonumber \\ 
    + \bigg[(cx)^2 - 2 csx + s& + \scA{s}{\ell m}{c}   \\ 
      & - \frac{(m+sx)^2}{1-x^2}\bigg]\swS{s}{\ell{m}}{x}{c} = 0,
\nonumber
\end{align}
where $c\ (=a\omega)$ is the oblateness parameter and $m$ the azimuthal separation constant.  Finally, $\ell$ is the harmonic mode index which labels the elements in the set of eigensolutions of the angular equation for fixed values of $s$, $m$, and $c$.

The separation constant for solutions to Eq.~(\ref{eqn:swSF_DiffEqn}) obey the following useful identities:\cite{cook-zalutskiy-2014}
\begin{subequations}\label{eqn:swSF_all_ident}
\begin{align}
\label{eqn:swSF_sA_ident}
\scA{-s}{\ell{m}}{c} &= \scA{s}{\ell{m}}{c} + 2s, \\
\label{eqn:swSF_mcA_ident}
\scA{s}{\ell(-m)}{c} &= \scA{s}{\ell{m}}{-c}, \\
\label{eqn:swSF_cA_ident}
\scA[*]{s}{\ell{m}}{c} &= \scA{s}{\ell{m}}{c^*}.
\end{align}
\end{subequations}
Together, Eqs.~(\ref{eqn:swSF_mcA_ident}) and (\ref{eqn:swSF_cA_ident}) yield $\scA{s}{\ell(-m)}{c}=\scA[*]{s}{\ell m}{-c^*}$.  Because of this, if $\omega_{\ell m}$ is an eigenvalue of Eq.~(\ref{eqn:radialR:Diff_Eqn}) with eigenfunction $R_{\ell m}(r)$, then $-\omega_{\ell(-m)}^*$ is also an eigenvalue with eigenfunction $R^*_{\ell(-m)}(r)$.  Because of this symmetry, modes of the Kerr geometry are seen to come in two related families of solutions.  These two solutions are often referred to as mirror solutions since the relation
\begin{align}\label{eqn:mirror mode symmetry}
   \omega^-_{\ell m} = -(\omega^+_{\ell(-m)})^*
 \end{align}
 represents reflection through the imaginary axis.  One family of solutions is labeled by $\omega^+_{\ell m}$ and the other by $\omega^-_{\ell m}$.  In general, the $\omega^+_{\ell m}$ modes are chosen to have a positive real component.  However, this choice is not always possible. For example, in the original family of TTMs, if we consider that $\omega^+_{\ell m}$ modes with $m<0$ exist, then they have a negative real component for $a>0$.  Because of Eq.~(\ref{eqn:mirror mode symmetry}), it is only necessary to determine modes in one of the two families.  Unless explicitly stated, we will work exclusively with the $\omega^+$ family of solutions and drop the $+$ superscript to simplify notation.

Both Eqs.~(\ref{eqn:radialR:Diff_Eqn}) and (\ref{eqn:swSF_DiffEqn}) are examples of the general class of confluent Heun equations.  We have found it particularly useful to work with the radial equation, Eq.~(\ref{eqn:radialR:Diff_Eqn}), within the context of confluent Heun theory\cite{Heun-eqn}.  The TTMs of the Kerr geometry can be obtained by solving a continued fraction equation to obtain a confluent Heun function and associated eigenvalue.  This method is often referred to as Leaver's method\cite{leaver-1985,cook-zalutskiy-2014} and is a frequently used method for obtaining QNMs.  On the other hand, TTMs also exist as confluent Heun polynomial solutions to the Teukolsky radial equation.  The derivation of the polynomial TTM modes can be found in Sec.~III.B of Ref.\cite{cook-zalutskiy-2014}.  There, we find that the condition for the existence of confluent Heun polynomial solutions is equivalent to the vanishing of the Starobinsky constant\cite{wald-1973,chandra-1984}.  We write the Starobinsky constant as\footnote{Note that $\lambdabar$ in Eq.~(\ref{eqn:Starobinsky_const}) is the same as Eq.~(\ref{eqn:lambdabar def}) for $s=-2$, but not for $s=2$.  This is an unfortunate notational carryover from Ref.~\cite{chandra-1984}.}
\begin{align}\label{eqn:Starobinsky_const}
  |\mathcal{Q}|^2 =\lambdabar^2(\lambdabar+2)^2 
  &+ 8\lambdabar\bar{a}\bar\omega\left(
      6(\bar{a}\bar\omega+m) -5\lambdabar(\bar{a}\bar\omega-m)\right)
      \nonumber\\ \mbox{}&
      + 144\bar\omega^2\left(1+\bar{a}^2(\bar{a}\bar\omega-m)^2\right),
\end{align}
where $\bar\omega\equiv M\omega$ is the dimensionless mode frequency. For TTM${}_{\rm L}$s, take $s=-2$ and
\begin{equation}
  \lambdabar=\lambdabar_-\equiv \scA{-2}{\ell{m}}{\bar{a}\bar\omega}
     + \bar{a}^2\bar\omega^2 - 2m\bar{a}\bar\omega.
\end{equation}
For TTM${}_{\rm R}$s, take $s=+2$ and
\begin{equation}
  \lambdabar=\lambdabar_+\equiv \scA{2}{\ell{m}}{\bar{a}\bar\omega}
     + \bar{a}^2\bar\omega^2 - 2m\bar{a}\bar\omega+4.
\end{equation}
However, because of Eq.~(\ref{eqn:swSF_sA_ident}), it follows that $\lambdabar_+=\lambdabar_-$ and we find that the TTM${}_{\rm L}$ and TTM${}_{\rm R}$ algebraically special modes\footnote{Polynomial solutions to the radial equation satisfying the vanishing of the Starobinsky constant are often referred to as being algebraically special\cite{chandra-1984}.} will share the same frequency spectrum.  

\subsection{Numerical solutions}
\label{sec:numerical methods}

For $0\le\bar{a}\le1$, numerical values for the TTM mode frequencies $\bar\omega$ are obtained by finding roots of Eq.~(\ref{eqn:Starobinsky_const}) where values for $\scA{\pm2}{\ell{m}}{\bar{a}\bar\omega}$ are obtained by solving the angular Teukolsky equation, Eq.~(\ref{eqn:swSF_DiffEqn}), using the spectral solver described in Sec.~II.D.1 of Ref.\cite{cook-zalutskiy-2014}.  Note that this is inherently an iterative approach since evaluation of $\scA{\pm2}{\ell m}{\bar{a}\bar\omega}$ requires prior knowledge of $\bar\omega$.

Initial guesses for $\bar\omega$ are obtained by plotting the ${\rm Re}[|{\mathcal Q}|^2]$ and ${\rm Im}[|{\mathcal Q}|^2]$ over some region of the complex plane covered by $\bar\omega$.  For fixed values of $m$, $s=\pm2$, and $\bar{a}$, the angular Teukolsky equation(\ref{eqn:swSF_DiffEqn}) is solved at each value of $\bar\omega$.  This results in a set of eigensolutions that we index by $\ell\ge\max(|m|,|s|)$.  We choose a specific value of $\ell$ and use that value of $\scA{\pm2}{\ell{m}}{\bar{a}\bar\omega}$ to evaluate $|{\mathcal Q}|^2$ at each value of $\bar\omega$.  Locations where both ${\rm Re}[|{\mathcal Q}|^2]=0$ and ${\rm Im}[|{\mathcal Q}|^2]=0$ represent solutions for $\bar\omega_{\ell mn}$.  The index $n$ is used to differentiate multiple solutions with the same values of $\ell$ and $m$.  For QNMs, the index $n$ is referred to as an overtone index.  For TTMs, it is used to differentiate families of solutions, but for consistency we will continue to refer to it as the overtone index.

Given a single converged solution $\omega_{\ell mn}$ at one value of $\bar{a}$, we construct a full sequence of solutions for a large set of values of $\bar{a}$ in the range $0\le\bar{a}\le1$.  Each solution in the sequence requires a choice, during the solution iteration, of possible separation constants from the set of possible eigensolutions to the angular equation for fixed $m$, $s$, and $c=\bar{a}\bar\omega$.  This choice is fixed by demanding, whenever possible, that the sequences of mode frequency $\bar\omega_{\ell mn}(\bar{a})$ and separation constant $\scA{s}{\ell m}{\bar{a}\bar\omega_{\ell mn}}$ are $C^1$ smooth.  However, this choice is not always possible.  For TTMs, as mentioned in the introduction, there appears to be a countably infinite set of points along two families of $m=0$ sequences at which the sequence is continuous, but not smooth.  Aspects of this behavior have already been observed\cite{onozawa-1997,cook-et-al-2018}.  Furthermore, for QNMs, there are even sequences along which the mode frequency intersects the negative imaginary axis and the sequences are not even continuous\cite{cook-zalutskiy-2016b}.  In order to conveniently handle situations where a sequence is not smooth, or not even continuous, we use the notation of ``overtone multiplets''\cite{cook-zalutskiy-2016b} to label each continuous segment of a sequence with the same overtone value but with an additional subscript on the overtone value\footnote{There are even situations when two seemingly separate sequences should logically have the same overtone (see Fig.~9 and associated text from Ref.~\cite{cook-zalutskiy-2016b}), and we also use overtone multiplets to label these sequences.}

\subsection{Asymptotic expansions}
\label{sec:asymptotic methods}
In addition to finding numerical sequences of TTMs parameterized by $\bar{a}$, we will also be interested in finding analytic asymptotic expansions for our solutions in the Schwarzschild limit $\bar{a}\to0$.  As we will show below, the two new families of TTMs we have found have Schwarzschild limits in which the mode frequencies approach complex infinity.  In this limit, the mode frequencies can be expressed in an asymptotic expansion as
\begin{align}\label{eqn:omega fit function}
  M\omega_{\ell mn}(\bar{a}) &= \sum_{p=0}^\infty{\frac{B_p(\ell,m,n)}{\bar{a}^{(4-p)/3}}}.
\end{align}
The separation constant is naturally considered as a function of the oblateness parameter $c=a\omega=\bar{a}\bar{\omega}$.  Its asymptotic expansion in terms of $c$ is in the limit as $c$ goes to complex infinity, but in terms of $\bar{a}$ is in the limit as $\bar{a}\to0$.  These two asymptotic expansions can be expressed as
\begin{align}\label{eqn:A vs c fit}
  \scA{s}{\ell m}{c} &= \sum_{p=0}^\infty{C_p(\ell,m,s) c^{1-p}} \\
\intertext{and}
\label{eqn:A vs a fit}
  \scA{s}{\ell m}{a\omega_{\ell mn}} &= \sum_{p=0}^\infty{D_p(\ell,m,n,s) \bar{a}^{(p-1)/3}}.
\end{align}
Clearly, the expansion coefficients in Eq.~(\ref{eqn:A vs a fit}) can be expressed in terms of those of Eqs.~(\ref{eqn:omega fit function}) and (\ref{eqn:A vs c fit}), where we find
\begin{subequations}\label{eqn:BCD relations all}
\begin{align}
\label{eqn:BCD0}
  D_0 &= B_0 C_0, \\
\label{eqn:BCD1}
  D_1 &= B_1 C_0 + C_1 ,\\
\label{eqn:BCD2}
  D_2 &= B_2 C_0 + \frac{C_2}{B_0}, \\
\label{eqn:BCD3}
  D_3 &= B_3 C_0 - \frac{B_1 C_2}{B_0^2} + \frac{C_3}{B_0^2}, \\
  &\vdots. \nonumber
\end{align}
\end{subequations}

If we insert Eqs.~(\ref{eqn:omega fit function}) and (\ref{eqn:A vs a fit}) into the Starobinsky constant Eq.~(\ref{eqn:Starobinsky_const}), we can expand it in powers of $\bar{a}$.  Since the existence of TTM solutions is equivalent to the vanishing of the Starobinsky constant, the vanishing of each term in the expansion can be used to fix the $B_p(\ell,m,n)$ coefficients.  Interestingly, as we will show below, simply fixing the leading order behavior in Eqs.~(\ref{eqn:omega fit function}) and (\ref{eqn:A vs c fit}) to their given values is sufficient to determine a set of possible values for $B_0$.

If the coefficients $C_p(\ell,m,s)$ in Eq.~(\ref{eqn:A vs c fit}) are known, then all of the $B_p(\ell,m,n)$ coefficients can be determined.  Unfortunately, a general analytic asymptotic expansion for the separation constant $\scA{s}{\ell m}{c}$ does not exist.  An asymptotic expansion has been determined analytically for oblate solutions where $c$ is purely real\cite{Casalas-oblate-2005}.  And, recently, an expansion in the prolate limit where $c$ is purely imaginary has been constructed based on numerical solutions\cite{VickersCook2022}.

In the absence of a known general asymptotic expansion for the separation constant $\scA{s}{\ell m}{c}$, we can attempt to determine the expansion coefficients $C_p(\ell,m,s)$ numerically in the asymptotic limit along our sequences of numerically generated solutions.  Since our numerical solution sequences contain both the mode frequency $\bar\omega_{\ell mn}(\bar{a})$ and the separation constant $\scA{\pm2}{\ell mn}{\bar{a}}$\footnote{Note the extra index $n$ on the separation constant is to denote the overtone family of the solution when the separation constant is parameterized by $\bar{a}$ instead of by $c=a\omega_{\ell mn}$.} parameterized by $\bar{a}$, there are two independent ways to determine the $C_p(\ell,m,s)$ coefficients.  Both approaches are based on the relationships between the $B_n$, $C_n$, and $D_n$ coefficients given in Eq.~(\ref{eqn:BCD relations all}) and by the fact that we can determine $B_0$ independently from the separation constant.

The first of the two approaches is based on fitting to the separation constant $\scA{s}{\ell mn}{\bar{a}}$ in a set of numerically generated solutions.  For each value of $\ell$ and $m$ for fixed $s$ and $n$, we fit for the coefficients $D_p(\ell,m,n,s)$.  Using Eq.~(\ref{eqn:BCD0}), we can determine $C_0(\ell,m,s)$ and then fit these to functions of $\ell$ and $m$ to obtain an analytic form for $C_0(\ell,m,s)$.  Examining the expansion of the Starobinsky constant, Eq.~(\ref{eqn:Starobinsky_const}), it turns out that knowledge of $B_p$ and $C_p$ for $0\le p<n$ is sufficient to determine $B_n$, so we also know $B_1$.  Using a greedy approach, we can consider $B_0$, $B_1$, and  $C_0$ as known and fit again for the coefficients $D_p(\ell,m,n,s)$.  Now using Eq.~(\ref{eqn:BCD1}), we can determine $C_1$ and $B_2$.  We can continue this process to sequentially extract values for $C_n$ and $B_{n+1}$ until we exhaust the numerical precision in our data.  We will refer to this approach as {\em angular fitting} since it involves fitting directly to the separation constant obtained from the angular equation(\ref{eqn:swSF_DiffEqn}).

The second approach is based on fitting to the complex mode frequencies $\bar\omega_{\ell mn}(\bar{a})$ in a set of numerically generated solutions.  This approach is similar to the first, except that the expansion of the Starobinsky constant, Eq.~(\ref{eqn:Starobinsky_const}), is used to express the expansion coefficients $B_n(\ell,m,n)$ as functions of $B_p(\ell,m,n)$ and $C_p(\ell,m,s)$ for $0\le p<n$.  We note that determining $B_1(\ell,m,n)$\footnote{Recall that $B_0$ is determined without fitting.} does require fitting Eq.~(\ref{eqn:A vs a fit}) to the separation constant data to determine $D_0$ and then obtain $C_0$ using the angular fitting approach.  However, all subsequent terms can be determined by an independent greedy approach.  With known values for $B_0$ and $B_1$, and with $B_2$ expressed as a function of $C_1$, we can fit Eq.~(\ref{eqn:omega fit function}) to the numerical data for $\bar\omega_{\ell mn}(\bar{a})$ to determine $C_1(\ell,m,s)$ and then fit these to functions of $\ell$ and $m$ to obtain an analytic form for $C_1(\ell,m,s)$.  We can continue this process to sequentially extract values for $C_{n-1}$ and $B_n$ until we exhaust the numerical precision in our data.  We will refer to this approach as {\em radial fitting} since it involves fitting directly to the mode frequency obtained from the radial equation(\ref{eqn:radialR:Diff_Eqn}).

\section{Numerical results}
\label{sec:numerical_results}

Numerical solutions have been constructed for TTMs for all three families mentioned in the introduction for all values of $m$ allowed for values of $2\le\ell\le8$.  Even though the TTM${}_{\rm L}$ and TTM${}_{\rm R}$ sequences share the same mode frequencies $\bar\omega_{\ell mn}(\bar{a})$, and the angular separation constants $\scA{\pm2}{\ell m}{\bar{a}\bar\omega_{\ell m n}}$ are related by Eq.~(\ref{eqn:swSF_sA_ident}), we have independently constructed both sequences and verified that the solutions are in agreement within numerical error.  Each mode sequence has been constructed for $0\le\bar{a}\le1$ with a maximum step size in $\bar{a}$ of $1/1000$, however, smaller step sizes were used when needed to ensure that the sequences appear smooth and deal appropriately with nonsmooth behavior.

\subsection{$\ell=2$ modes}
\label{sec:l2 plots}

\begin{figure}
    \centering
  \includegraphics[width=\linewidth,clip]{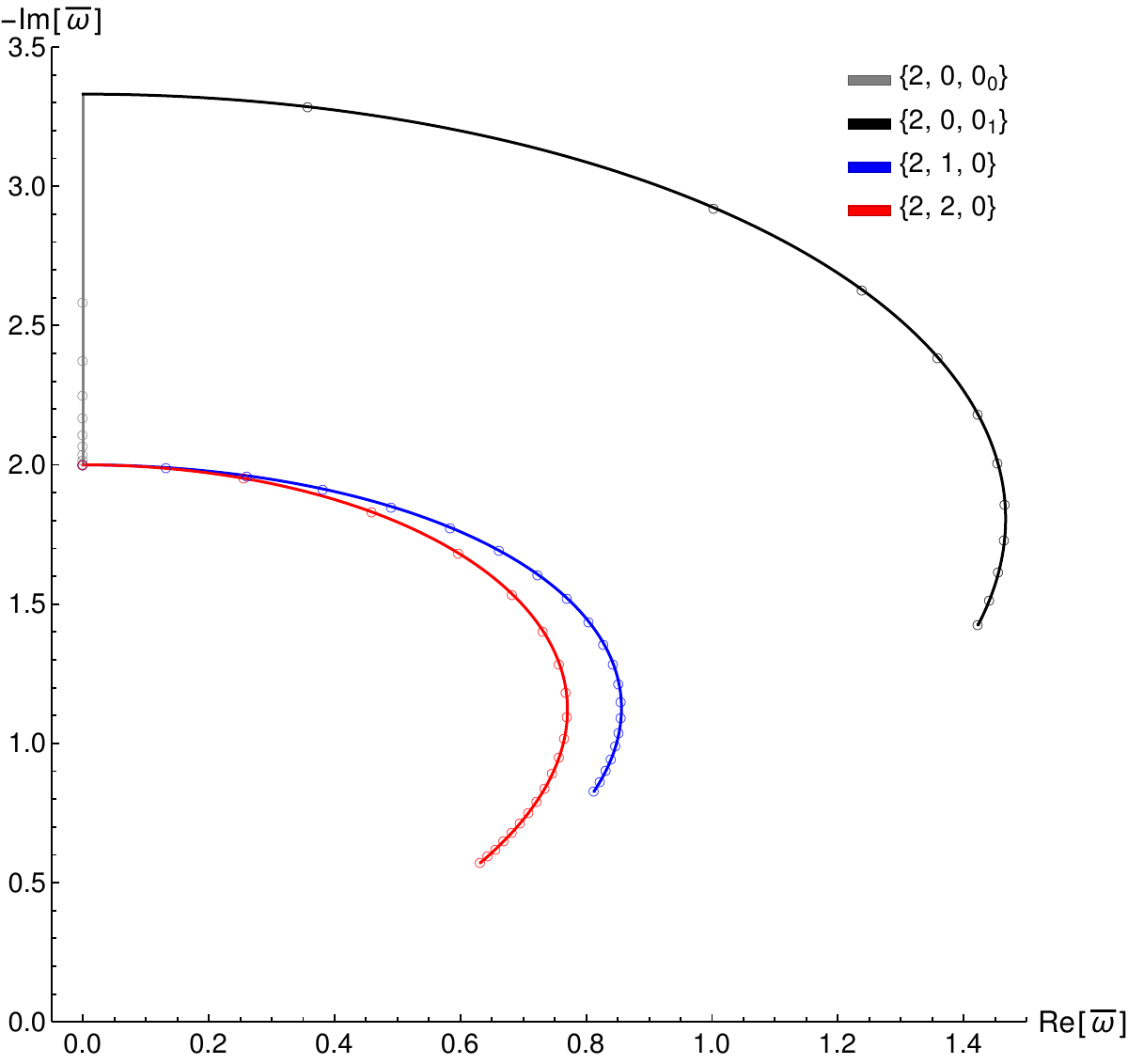}
    \caption{Kerr TTM mode sequences for $\ell=2$ in the original family denoted by $n=0$.  The $m=0$ sequence is split into two smooth segments distinguished by the use of overtone-multiplet notation.  One segment is labeled as $\{2,0,0_0\}$ and the other by $\{2,0,0_1\}$, and are drawn in the figure by gray and black lines respectively.  The $m=1$ sequence is drawn as a blue line, and the $m=2$ sequence is drawn as a red line.  Along each sequence are open circles drawn at values of $\bar{a}$ that are multiples of $0.05$.}
    \label{fig:l2 n0 all m}
\end{figure}

We will begin exploring the new TTMs by examining the simplest cases with $\ell=2$.  Detailed plots of the original $\ell=2$ TTM sequences have been presented previously, first by Onozawa\cite{onozawa-1997}, and subsequently in Refs.~\cite{cook-zalutskiy-2014,cook-et-al-2018}.  We reproduce this figure from our current data sets in Fig.~\ref{fig:l2 n0 all m}.
Because of the degeneracy of the mirror modes, $\bar\omega^-_{\ell m0}=\bar\omega^+_{\ell m0}$, for the original($n=0$) family of modes, we only plot sequences for $\bar\omega_{\ell m0}$ for $m\ge0$.  All three sequences begin at the Schwarzschild limit($\bar{a}=0$) which has a frequency of $\bar\omega_{2m0}=-2i$ as given by Eq.~(\ref{eqn:alg-spec-sch}).  The mode frequencies for sequences with $m\ne0$ immediately take on complex values for $\bar{a}>0$, however the $m=0$ sequence initially moves along the negative imaginary axis (NIA) before discontinuously changing direction.  At a value of $\bar{a}\equiv\hat{a}_2\approx0.4944459549145$ and $\bar\omega_{200}\approx-3.330810232354i$, the mode frequencies along the sequence abruptly begin to take on fully complex values.  Because of the nonsmooth behavior of the $m=0$ sequence, it is separated into two overtone-multiplet sequences, each of which is smooth over its domain.  The sequence that smoothly connects to the Schwarzschild limit is denoted by $n=0_0$, while the sequence which smoothly connects to the maximally rotating limit with $\bar{a}=1$ is denoted by $n=0_1$.

\begin{figure}
    \centering
  \includegraphics[width=\linewidth,clip]{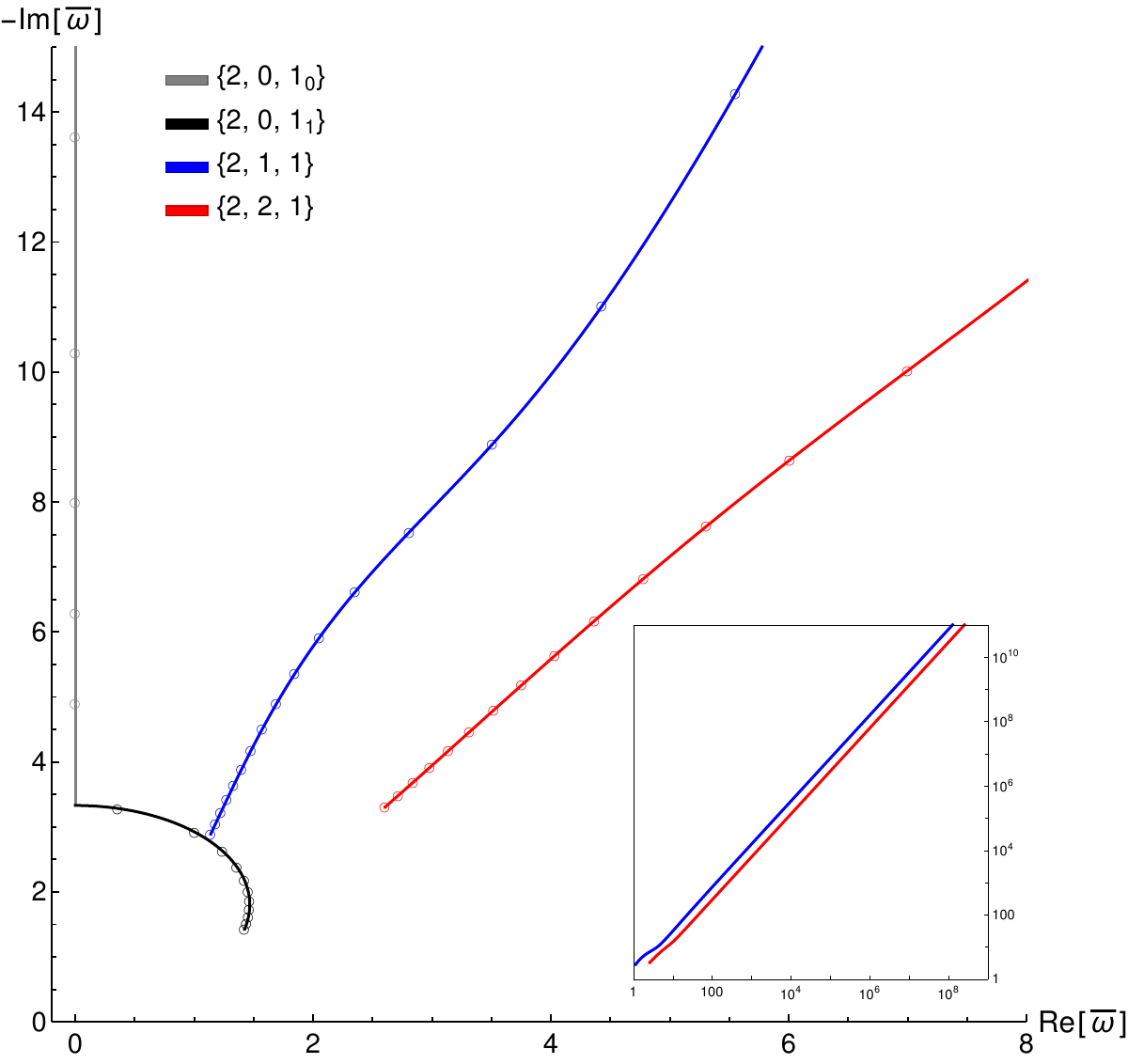}
    \caption{Kerr TTM mode sequences for $\ell=2$ in the first new family denoted by $n=1$.  See the caption to Fig.~\ref{fig:l2 n0 all m} for additional details.  The main plot shows a detailed view of the behavior of the modes for large values of $\bar{a}$.  The inset highlights the asymptotic behavior of the $m>0$ modes by using logarithmic scaling on both axes.  Note that the Schwarzschild limit occurs at complex infinity.}
    \label{fig:l2 n1 all m}
\end{figure}
In Fig.~\ref{fig:l2 n0 all m}, the first segment along the $m=0$ sequence is plotted as a gray line along the NIA.  Note that the imaginary axis has been shifted to the left of the origin so that this line segment can be easily seen.  Each sequence is labeled by its values of $\{\ell,m,n\}$.  The second segment, labeled as $\{2,0,0_1\}$, is plotted as a black line extending through the complex $\bar\omega$ plane.  Note that we follow a common convention of flipping the imaginary axis for plots of Kerr-mode frequencies.  Negative imaginary values increase in magnitude as we move up in the plot, and we recall that negative imaginary mode frequencies correspond to stable, exponentially damped modes.  Finally, we note that along each sequence there are a number of small open circles.  These circles are separated by a stride of $0.05$ in $\bar{a}$ and are intended to give a sense of how $\bar{a}$ changes along each sequence.

The $\ell=2$ frequencies for modes in the first of the two new families of TTMs are plotted in Fig.~\ref{fig:l2 n1 all m}.  Sequences in this family are designated by an overtone index of $n=1$.  As with the $n=0$ family, the modes in this family are degenerate with their mirror modes, $\bar\omega^-_{\ell m1}=\bar\omega^+_{\ell m1}$, and we only plot sequences for $\bar\omega_{\ell m1}$ for $m\ge0$.  In contrast to the $n=0$ family of modes, the mode frequencies in the Schwarzschild limit are not finite, but exist at complex infinity.  The main plot in Fig.~\ref{fig:l2 n1 all m} shows a detailed view of the behavior of the modes for large values of $\bar{a}$.  For each sequence in this portion of the figure, the end of the sequence is the maximally rotating limit of $\bar{a}=1$, and the sequence extends back to the Schwarzschild limit of $\bar{a}=0$ as we move upward along the sequence.  The smaller inset plot in the figure highlights the asymptotic behavior of the $m>0$ modes by using logarithmic scaling along both axes.  The asymptotic(Schwarzschild limit) portion of each sequence is linear in the log-log plot with a slope of approximately $4/3$, but each sequence seems to have a unique intercept.

The $m=0$ sequence in Fig.~\ref{fig:l2 n1 all m} clearly shows different behavior from the sequences with $m\ne0$.  As with the $m=0$ sequence in the $n=0$ family, it has a segment that is connected to the Schwarzschild limit and along which the mode frequency lies entirely along the NIA.  This segment is labeled as $\{2,0,1_0\}$ and is drawn as a gray line in the figure.  At $\bar{a}=\hat{a}_2\approx0.4944459549145$ and $\bar\omega_{201}=\hat\omega_2\approx-3.330810232354i$, the sequence abruptly changes direction and takes on complex values.  The second segment of the sequence is labeled as $\{2,0,1_1\}$ and is drawn as a black line in the figure.  The $m=0$ sequence for $\ell=2$, and in fact for each value of $\ell$, in the $n=1$ family was first discovered in Ref.~\cite{cook-et-al-2018}.  In that work, the $\{2,0,1_0\}$ segment was considered as a third branch of the $\ell=2$, $m=0$ TTM sequence.  We claim that this segment actually belongs to the $n=1$ family of TTMs.  This may seem strange since the slope of this segment is clearly zero, apparently differing from the behavior of the modes with $m\ne0$.  However, we will fully justify this claim in Sec.~\ref{sec:asymptotic behavior}.  For now, we note that the $\{2,0,1_1\}$ segment is degenerate with the $\{2,0,0_1\}$ segment from the original $n=0$ family.

\begin{figure}
    \centering
  \includegraphics[width=\linewidth,clip]{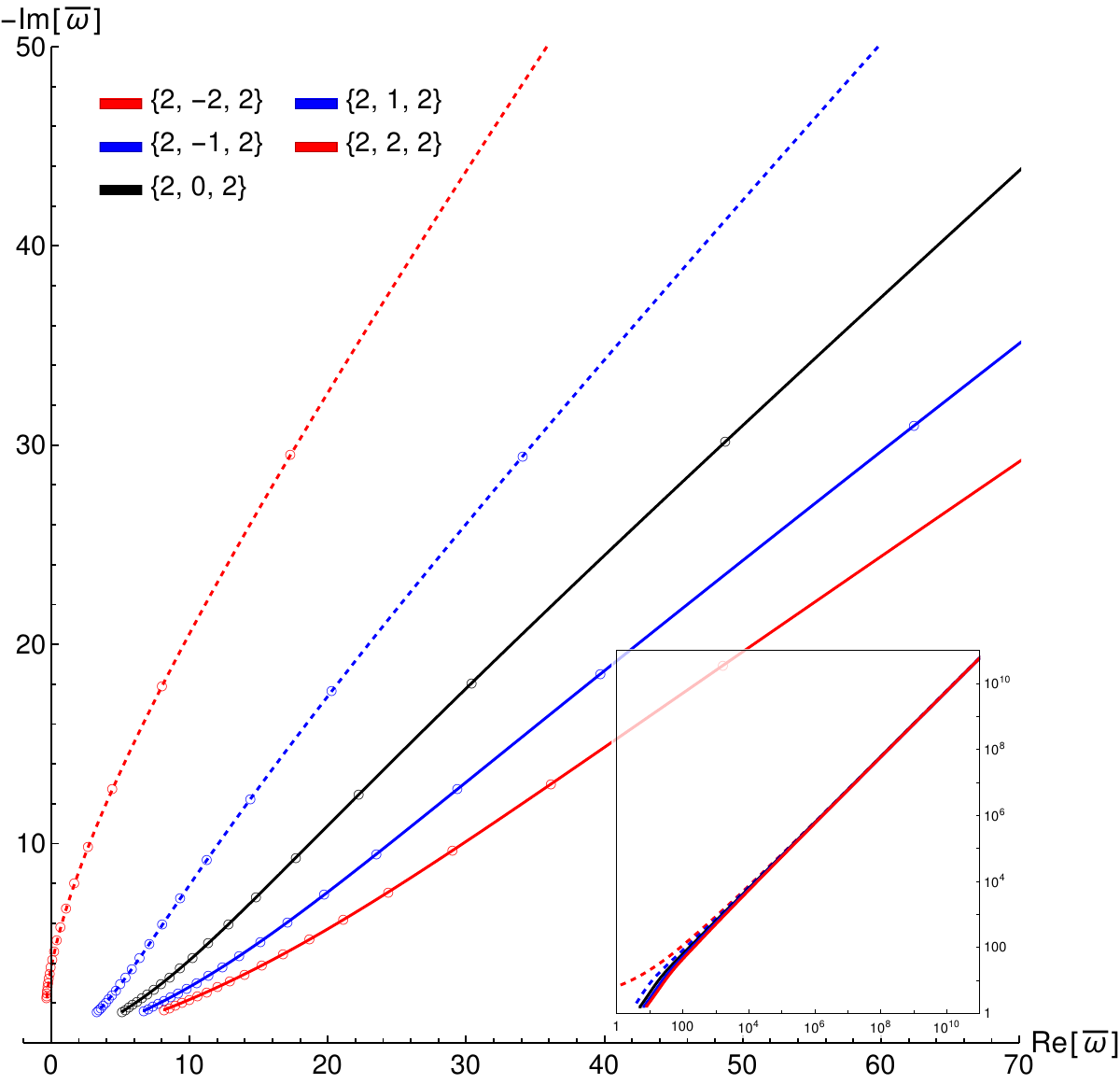}
    \caption{Kerr TTM mode sequences for $\ell=2$ in the second new family denoted by $n=2$.  See the caption to Fig.~\ref{fig:l2 n0 all m} for additional details.  The main plot shows a detailed view of the behavior of the modes for large values of $\bar{a}$.  Mode sequences with negative values of $m$ are drawn as dashed lines.  The inset highlights the asymptotic behavior of the modes by using logarithmic scaling on both axes.  Note that the Schwarzschild limit occurs at complex infinity.}
    \label{fig:l2 n2 all m}
\end{figure}

Finally, the $\ell=2$ frequencies for modes in the second of the two new families of TTMs are plotted in Fig.~\ref{fig:l2 n2 all m}.  Sequences in this family are designated by an overtone index of $n=2$.  Unlike the modes in the previous two families, the modes in this family are not degenerate with their mirror modes, and we must plot all allowed values of $m$ in order to represent all sequences which are unique up to reflection across the NIA.  
And, there are several other aspects of this family of modes which are unique.  The $m=0$ sequence is no longer purely imaginary and it is smooth over its entire extent.  For $\ell=2$, the $\{2,-2,2\}$ mode frequencies cross the NIA as $\bar{a}$ approaches the extreme rotation limit so that the real part of the frequency can take on negative values.  In general, we define the $\omega^+$ set of modes to have a positive real frequency component.  But, as mentioned in Sec.~\ref{sec:methods} this is not always possible.  The only alternative allowing $\rm{Re}[\omega^+]$ to maintain positive frequency would be to discontinuously stop an $\omega^+_{\ell m2}$ sequence as it crosses the NIA and relabel its continuation as $\omega^-_{\ell(-m)2}$.  This seems to be a drastic and undesirable solution.  As with the $n=1$ family of modes, in the Schwarzschild limit, the mode frequencies of the $n=2$ family also exist at complex infinity.  The smaller inset plot in Fig.~\ref{fig:l2 n2 all m} highlights the asymptotic behavior of the modes by using logarithmic scaling along both axes.  The asymptotic(Schwarzschild limit) portion of each sequence is linear in the log-log plot with a slope of approximately $1$, and each sequence seems to have a common intercept.

\subsection{General behavior of all modes}
\label{sec:behavior of all modes}

\begin{figure}
    \centering
  \includegraphics[width=\linewidth,clip]{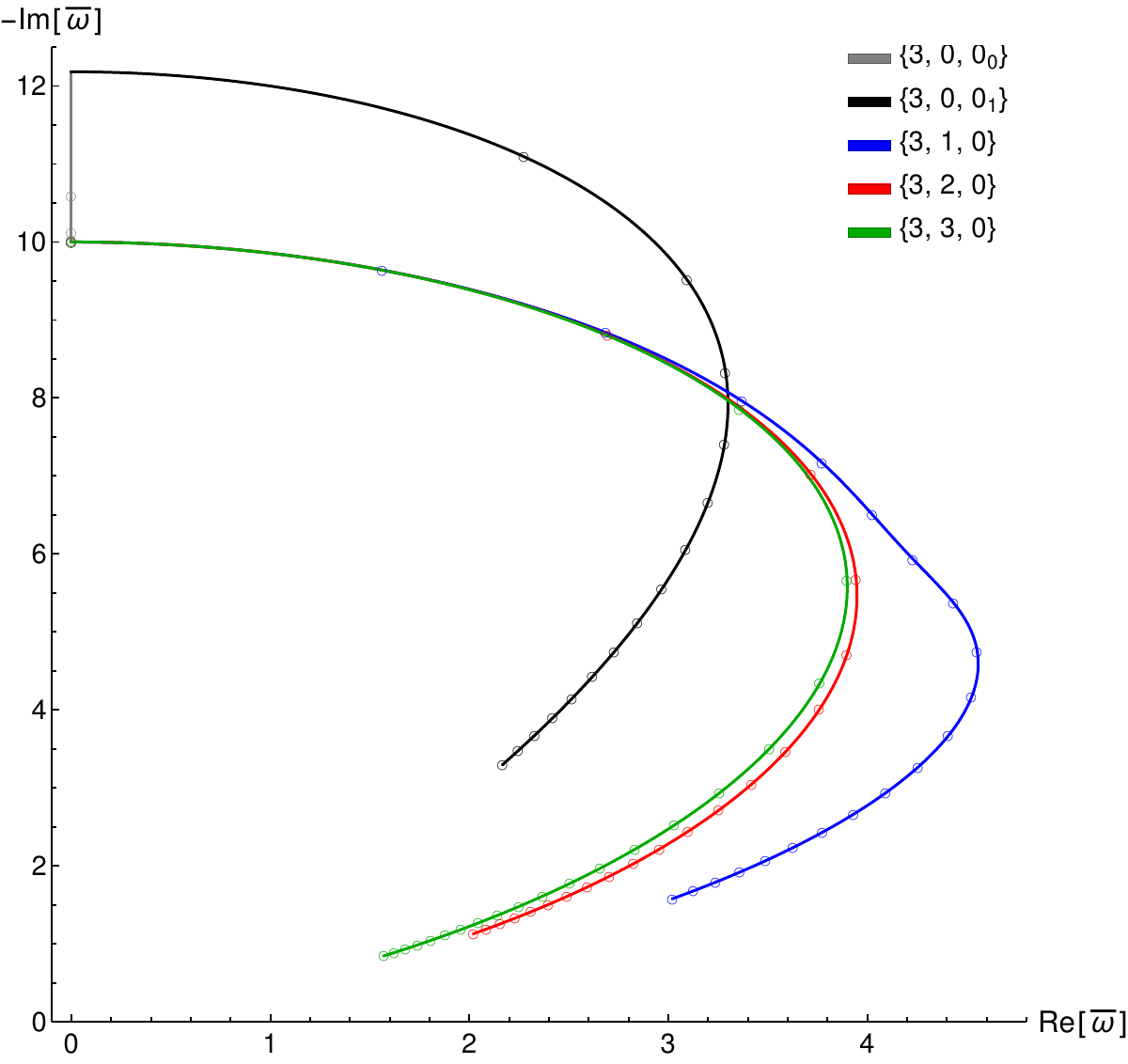}
    \caption{Kerr TTM mode sequences for $\ell=3$ in the original family denoted by $n=0$.  See the caption to Fig.~\ref{fig:l2 n0 all m} for additional details.}
    \label{fig:l3 n0 all m}
\end{figure}

\begin{figure}
    \centering
  \includegraphics[width=\linewidth,clip]{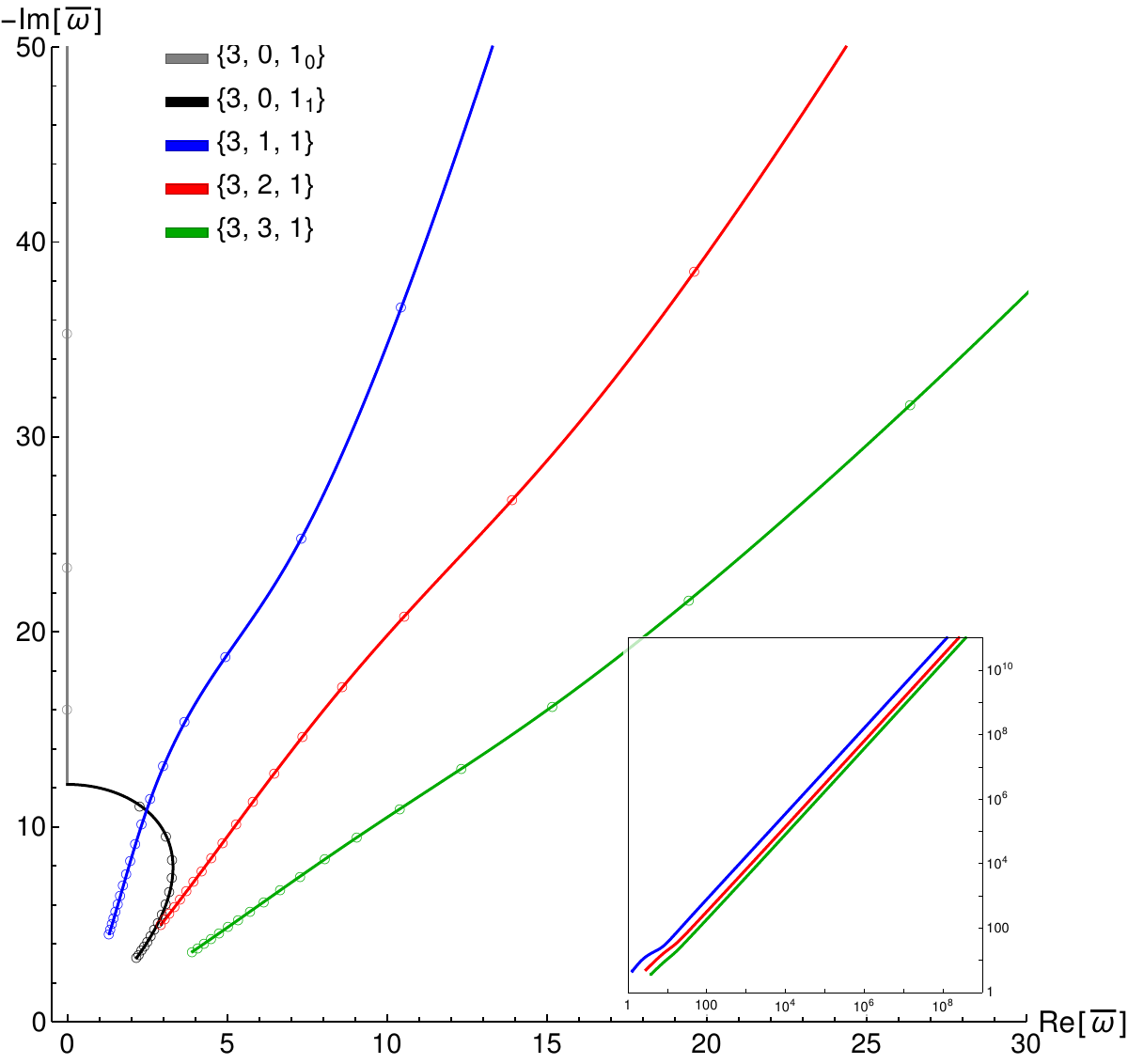}
    \caption{Kerr TTM mode sequences for $\ell=3$ in the first new family denoted by $n=1$.  See the caption to Figs.~\ref{fig:l2 n0 all m} and \ref{fig:l2 n1 all m} for additional details.}
    \label{fig:l3 n1 all m}
\end{figure}

\begin{figure}
    \centering
  \includegraphics[width=\linewidth,clip]{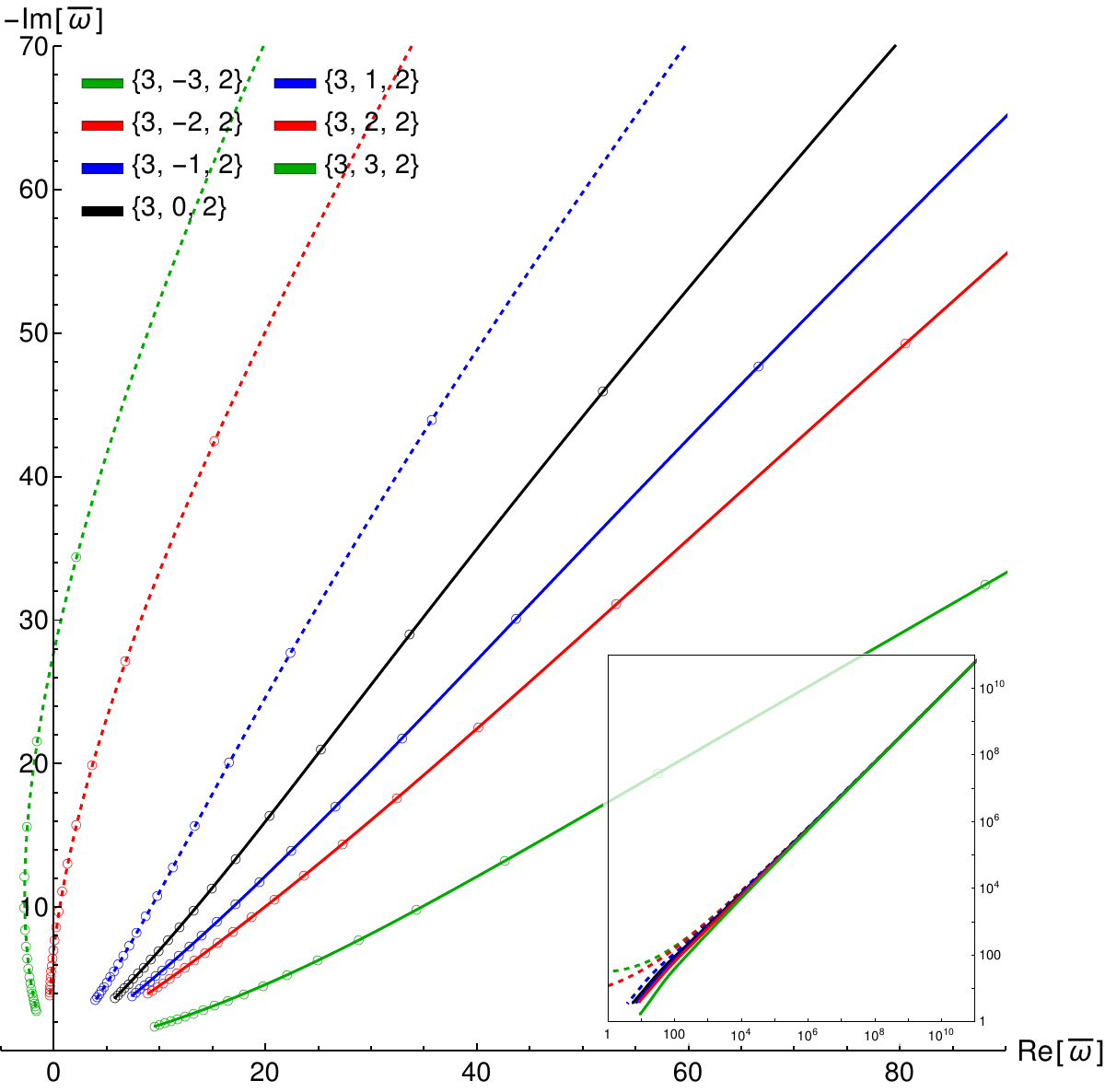}
    \caption{Kerr TTM mode sequences for $\ell=3$ in the second new family denoted by $n=2$.  See the caption to Fig.~\ref{fig:l2 n0 all m} and \ref{fig:l2 n2 all m} for additional details.}
    \label{fig:l3 n2 all m}
\end{figure}

We have constructed full sequences for all modes with $\ell\le8$.  For $\ell>2$, the observed behavior of all three families of modes is very similar to that of the three $\ell=2$ families.   Figures~\ref{fig:l3 n0 all m}---\ref{fig:l3 n2 all m} present mode frequency sequences for the three $\ell=3$ families.
Plots of the mode frequencies for each of the three families of TTMs for $4\le\ell\le8$ can be found in Figs.~\ref{fig:l4 n012 all m}---\ref{fig:l8 n012 all m} in the Appendix~\ref{sec:appendix figures}.  All members of the original family of TTMs have a Schwarzschild limit that is at a finite imaginary frequency given by Eq.~(\ref{eqn:alg-spec-sch}).  The $m=0$ sequences of mode frequencies for this family always begin moving along the NIA, but not always in the same direction.  For $\ell<4$, the mode frequencies initially move in the direction of increasing values for $|\bar\omega_{\ell00_0}|$, but for $\ell\ge4$, mode frequencies initially move toward the origin of the NIA.  In all cases, the $\{\ell,0,0_0\}$ segments exhibit nonsmooth behavior at values of $\bar{a}=\hat{a}_\ell$ and mode frequencies $\bar\omega_{\ell00_0}=\hat{\omega}_\ell$ given in Table~\ref{table:hat a hat omega}.  Beyond this point the mode frequencies take on general complex values.  The next portion of each sequence of mode frequencies is again smooth and labeled by $\{\ell,0,0_1\}$.  The remaining sequences with $0<|m|\le\ell$ move immediately away from the NIA, and all of the modes are degenerate with their mirror modes so that $\omega^-_{\ell m0}=\omega^+_{\ell m0}$.
\begin{table}[h]
 \begin{tabular}{c| d{14} d{14}}
 \multicolumn{1}{c|}{$\ell$} & \multicolumn{1}{c}{$\hat{a}_\ell$} & \multicolumn{1}{c}{$\hat\omega_{\ell}$}  \\
\hline
2 & \ 0.4944459549145 &   -3.330810232354i \\
3 &   0.2731627005644 &  -12.18211379576i \\
4 &   0.1896558549391 &  -26.28539809866i \\
5 &   0.1455356200824 &  -45.47056768462i \\
6 &   0.1181734597855 &  -69.6786356417i  \\
7 &   0.0995177455293 &  -98.8818609388i  \\
8 &   0.0859721462803 & -133.0645162405i
\end{tabular}
\caption{Critical values of $\bar{a}$ and $\bar\omega$ at which the $m=0$ sequences of the $n=0$ and $n=1$ mode frequency sequences show $C^0$ behavior. Values of $\bar{a}$ between the end of the $\{\ell,0,0_0\}$ and the beginning of the $\{\ell,0,0_1\}$ differ by $10^{-3}\times2^{-35}$ which fixes the accuracy for the values for $\hat{a}_\ell$.  The values for $\hat\omega_\ell$ are obtained from the limit of the $\{\ell,0,0_1\}$ sequences as they approach $\hat{a}_\ell$.}
\label{table:hat a hat omega}
\end{table}

All members of the $n=1$ family of TTMs have a Schwarzschild limit at complex infinity. The $m=0$ sequences of mode frequencies for this family always move along the NIA toward $\hat\omega_\ell$, and at this point exhibit nonsmooth behavior and turn off of the NIA.  The initial smooth segment with $\bar{a}\le\hat{a}_\ell$ is labeled by $\{\ell,0,1_0\}$ and the second smooth segment has $\bar{a}>\hat{a}_\ell$.  Again, in all cases, the second segment, labeled by $\{\ell,0,1_1\}$, is degenerate with the $\{\ell,0,0_1\}$ segment.  As with all of the modes in the original $n=0$ family, all of the modes are degenerate with their mirror modes so that $\omega^-_{\ell m1}=\omega^+_{\ell m1}$.

Finally, all members of the $n=2$ family of TTMs also have a Schwarzschild limit at complex infinity. The $m=0$ sequences of mode frequencies for this family always take on fully complex values, and none of the modes are degenerate with their mirror modes.

\subsection{Asymptotic behavior}
\label{sec:asymptotic behavior}

The asymptotic behavior of the $n=1$ and $n=2$ families of TTMs can be cleanly extracted to some order directly from the numerical sequences we have plotted in Figs.~\ref{fig:l2 n1 all m}, \ref{fig:l2 n2 all m}, \ref{fig:l3 n1 all m}, \ref{fig:l3 n2 all m}, and similar plots in Appendix~\ref{sec:appendix figures}.  We use data from the first $40$ data points in each sequence, which cover a range of $6.25\times10^{-5}\ge\bar{a}>3.8\times10^{-9}$, and corresponding values of the oblateness parameter $c=\bar{a}\bar\omega$ in the range $100<|c|<1475$.

From log-log plots of the real and imaginary parts of both $\bar\omega(\bar{a})$ and $\scA{\pm2}{\ell mn}{\bar{a}}$, the leading order behavior assumed in Eqs.~(\ref{eqn:omega fit function}) and (\ref{eqn:A vs a fit}) can be immediately confirmed.  Furthermore, the $n=1$, $m>0$ sequences strongly suggest that the expansions increase and decrease, respectively, by factors of $\bar{a}^{1/3}$ motivating our ansatz for the form of these expansions.  It also justifies the leading order behavior in Eq.~(\ref{eqn:A vs c fit}).

Without knowing any of the expansion coefficients in Eqs.~(\ref{eqn:omega fit function}) and (\ref{eqn:A vs a fit}), we can insert these into the Starobinsky constant, (\ref{eqn:Starobinsky_const}) and expand the result in powers of $\bar{a}$.  Since the vanishing of the Starobinsky constant defines a TTM, each term in the expansion should vanish.  The leading order term at order $\mathcal{O}(\bar{a}^{-8/3})$ yields
\begin{align}
  B_0^2(144+B_0^6)=0.
\end{align}
$B_0=0$ is a double root that does not match our solutions. There remain three complex-conjugate pairs of solutions.  The solution $B_0=-i2^{2/3}3^{1/3}$ agrees with some of our solutions and corresponds to exponentially damped modes.  Its conjugate represents an exponentially growing mode and will be ignored.  The solutions $B_0=-(-1)^{5/6}2^{2/3}3^{1/3}$ and $B_0=(-1)^{1/6}2^{2/3}3^{1/3}$ represent mirror-mode solutions that agree with the remainder of our data and correspond to exponentially damped modes.  Again, their conjugates represent exponentially growing modes and will be ignored.  Of the two mirror-mode solutions, $B_0=-(-1)^{5/6}2^{2/3}3^{1/3}$ has a positive real component and we will exclusively fit data for this version.  We then have the leading order behavior for the two new TTM families:
\begin{equation}\label{eqn:omega leading order}
  M\omega_{\ell mn} = \left\{\begin{array}{rl}
    -\frac{i2^{2/3}3^{1/3}}{\bar{a}^{4/3}} + \mathcal{O}(\bar{a}^{-1}) &:\ n=1, \\
    -\frac{i(-1)^{1/3}2^{2/3}3^{1/3}}{\bar{a}^{4/3}} + \mathcal{O}(\bar{a}^{-1}) &:\ n=2,
  \end{array}\right.
\end{equation}
and we note that the leading order behavior of $\omega_{\ell m1}$ and $\omega_{\ell m2}$ differ simply by a phase rotation of $\pi/3$.

Using the angular fitting approach outlined in Sec.~\ref{sec:asymptotic methods}, we fit the data for $\scA{\pm2}{\ell mn}{\bar{a}}$ to obtain values for $D_0(\ell,m,n,s)$.  Using Eq.~(\ref{eqn:BCD0}) and the appropriate version of $B_0(\ell,m,n)$ for each family, we find that $C_0$ takes on the values of odd imaginary integers.  That is
\begin{align}\label{eqn:C0 value}
  C_0 = C_0(\ell,m,s)=\frac{D_0(\ell,m,n,s)}{B_0(\ell,m,n)}=i(2\bar{L}+1),
\end{align}
where we define $\bar{L}$ via
\begin{align}\label{eqn:bar L def}
  \bar{L} \equiv \ell - \max(|m|,|s|).
\end{align}
The form of $C_0$ is not too surprising.  It is consistent with the leading order asymptotic behavior for $\scA{s}{\ell m}{c}$ obtained in Ref.~\cite{berticardosocasals-2006} for purely imaginary values of $c$ referred to as the prolate case.  However, we note that we are considering general, complex values of $c$ in the asymptotic limit and agreement with the prolate limit was not guaranteed.  The parameter $\bar{L}$ is a non-negative integer and, in the prolate case, has been shown to be associated with the number of real zeros of the spin-weighted spheroidal function\cite{berticardosocasals-2006}.  We use an overbar on $L$ to emphasize that it is associated with the asymptotic limit, not because it is dimensionless.  This notation is prudent because the value of $L$ in the limit of small $c$, and in the asymptotic limit may not match\cite{VickersCook2022}.  Nevertheless, this observed behavior in $\scA{\pm2}{\ell m}{c}$ is sufficient to assign to each sequence a unique mode index $\ell$ through Eq.~(\ref{eqn:bar L def}).

With an analytic form for $C_0$ fixed we now have an analytic form for $D_0$
\begin{align}\label{eqn:D0 value}
  D_0 = D_0(\ell,m,B_0)=i(2\bar{L}+1)B_0.
\end{align}
And, from the next term in the expansion of the Starobinsky constant at $\mathcal{O}(\bar{a}^{-7/3})$, we find that
\begin{align}
  B_1 = B_1(\ell,m) = \frac23(2m - i(2\bar{L}+1))
\end{align}
for both the $n=1$ and $n=2$ families.

With the first two terms in Eq.~(\ref{eqn:omega fit function}) determined, we can now analytically confirm the slopes seen in the log-log plots in Figs.~\ref{fig:l2 n1 all m}, \ref{fig:l2 n2 all m}, \ref{fig:l3 n1 all m}, \ref{fig:l3 n2 all m}, and similar plots in Figs.~\ref{fig:l4 n012 all m}---\ref{fig:l8 n012 all m}.  Because $B_0$ is imaginary for $n=1$, we have that ${\rm Im}[\bar\omega_{\ell m1}]$ scales as $\bar{a}^{-4/3}$ while the ${\rm Re}[\bar\omega_{\ell m1}]$ scales as $\bar{a}^{-1}$ and depends on $m$.  This immediately leads to a slope of $4/3$ for the asymptotic portion of the sequences and an intercept that depends on $m$, exactly as observed.  For the $n=2$ family, $B_0$ is complex so both the real and imaginary parts of $\bar\omega_{\ell m2}$ scale as $\bar{a}^{-4/3}$, so the asymptotic portion of these sequences should have a slope of $1$ and a common intercept, exactly as observed.

The term at order $\mathcal{O}(\bar{a}^{-2})$ of the expansion of the Starobinsky constant is somewhat more complicated to deal with because it depends on the spin weight $s=\pm2$, and on the overtone, $n=1$ or $n=2$.  We have carried out the fitting for both TTM${}_{\rm L}$s with $s=-2$, and TTM${}_{\rm R}$s with $s=+2$, but will only outline the fitting details for the TTM${}_{\rm R}$s.  The fits depend on $s$ through the separation constant $\scAnorm{s}{\ell m}{c}$ and through the Starobinsky constant.  The overtone dependence, we recall, comes from the fact that our numerical solutions show two unique leading order behaviors for the complex mode frequencies as shown in Eq.~(\ref{eqn:omega leading order}).  So, we must separately fit our expansions to the $n=1$ and $n=2$ data sets.  As we will see below, most (but not all) of the $B_n$ expansion coefficients depend on $n$.  For TTM${}_{\rm R}$ data with $s=2$, we find\footnote{For TTM${}_{\rm L}$ data with $s=-2$, the number -29 in the numerators in Eq.~(\ref{eqn:B2}) is replaced by -53.}
\begin{align}\label{eqn:B2}
  B_2 &= B_2(\ell,m,n,C_1) \\
  &= \left\{\begin{array}{rl}
    -i\frac{4(\bar{L}+im)(\bar{L}+1+im)-29+6C_1}{9\times2^{2/3}3^{1/3}} &:\ n=1, \nonumber \\
    -i(-1)^{-1/3}\frac{4(\bar{L}+im)(\bar{L}+1+im)-29+6C_1}{9\times2^{2/3}3^{1/3}} &:\ n=2,
  \end{array}\right.
  \intertext{and}
  D_1 &= D_1(\ell,m,C_1)=\frac23(2\bar{L}+1)(2\bar{L}+1+2im)+C_1.
\end{align}
Fitting this to our data sets using either angular or radial fitting, we can determine $C_1$.  The results from the angular and radial methods, and for fitting to the $n=1$ and $n=2$ families, are nearly identical and we display the angular fitting results from the $n=1$ family in Table~\ref{table:C1}.  
\begin{table}
 \begin{tabular}{c| d{9} d{7}}
 &\multicolumn{1}{c}{Estimate} & \multicolumn{1}{c}{$\sigma$}  \\
\hline
 $1$ & 4.9999896 & 3.7\times10^{-6} \\
 $\bar{L}$ & -1.9999954 & 2.5\times10^{-6} \\
 $\bar{L}^2$ & -2.0000006 & 4.4\times10^{-7} \\
 $m^2$ & 4.0000008 & 1.0\times10^{-7}
\end{tabular}
\caption{Angular fit results from fitting $4C_1$ to the $n=1$ family of TTMs.  The first column lists the $\bar{L}$ and $m$ dependence of each term in the linear fit function.  The second column lists the coefficient for that term determined by the fit.  The third column lists the uncertainty estimate for the fit coefficient.  Each term is consistent with an integer, and the result of the fit is displayed in Eq.~(\ref{eqn:C1}).}
\label{table:C1}
\end{table}
While $C_1$ depends on $s$, remarkably we find that it is independent of $n$.  The result is
\begin{align}\label{eqn:C1}
   C_1(\ell,m,s) = -\frac14 \left[2 \bar{L} (\bar{L}+1) -4 m^2 -4 s(s-1)+3\right].
 \end{align}
Note that terms $-4s(s-1)+3$ cannot be uniquely determined by the $s=\pm2$ cases.  This expression agrees with the constant term in Table~\ref{table:C1} for the case of $s=2$, and also agrees with the result from fitting the $s=-2$ case.  Furthermore, this expression is consistent with Eq.~(\ref{eqn:swSF_sA_ident}) which fixes the coefficient in the term linear in $s$.  But the remaining constant and the coefficient of the $s^2$ term are not unique.  The form displayed in Eq.~(\ref{eqn:C1}) is chosen to agree with the asymptotic expansion for $\scA{s}{\ell m}{c}$ for the prolate case derived in Ref.~\cite{VickersCook2022} where the $s$ dependence was uniquely determined.

We can substitute Eq~(\ref{eqn:C1}) into Eq.~(\ref{eqn:B2}) to fully determine $B_2$.  Since we expect the TTM solutions to be isospectral, it is not surprising that the result [see Eq.~(\ref{eqn:Omega 2}) below] depends on $n$ but not $s=\pm2$.  We find this to be true for all of the $B_p$ coefficients.  We continued this procedure using both the angular and radial fitting methods outlined in Sec.~\ref{sec:asymptotic methods} on both the $n=1$ and $n=2$ families to determine the $C_2$, $C_3$, and $C_4$ coefficients in the asymptotic expansion for $\scA{\pm2}{\ell m}{c}$.  The fitting results from the angular fitting method applied to the $n=1$ family of data are displayed in Tables~\ref{table:C2}---\ref{table:C4}.  The real and imaginary parts of each coefficient were fit separately.  Again, all of the $C_p$ coefficients are independent of $n$.  The full expression for $\scA{s}{\ell m}{c}$ is shown in Eq.~(\ref{eqn:asymptotic normal solution}), and we will discuss this solution in more detail below.  Before doing so, we will finish the derivation of the asymptotic expansion for $\bar\omega$.

\begin{table}
 \begin{tabular}{c| d{10} d{7}}
 &\multicolumn{1}{c}{Estimate} & \multicolumn{1}{c}{$\sigma$}  \\
\hline
 $1$ & 2.059\times10^{-6} & 7.8\times10^{-7} \\
 m & 7.9999988 & 2.2\times10^{-7} \\
\hline
\hline
 $1$ & -60.999988 & 0.000010 \\
 $\bar{L}$ & -121.00001 & 0.000014 \\
 $\bar{L}^2$ & 3.0000069 & 5.8\times10^{-6} \\
 $\bar{L}^3$ & 1.9999992 & 6.5\times10^{-7} \\
 $m^2$ & -8.0000009 & 3.1\times10^{-7} \\
 $\bar{L} m^2$ & -16.000001 & 2.3\times10^{-7}
\end{tabular}
\caption{Angular fit results from fitting ${\rm Re}[C_2]$ (upper rows) and $16{\rm Im}[C_2]$ (lower rows) to the $n=1$ family of TTMs.  See the caption for Table~\ref{table:C1} for additional details.}
\label{table:C2}
\end{table}

\begin{table}
 \begin{tabular}{c| d{10} d{7}}
 &\multicolumn{1}{c}{Estimate} & \multicolumn{1}{c}{$\sigma$}  \\
\hline
 $1$ & 654.99986 & 0.000061 \\
 $\bar{L}$ & -732.99984 & 0.00015 \\
 $\bar{L}^2$ & -728.00010 & 0.00011 \\
 $\bar{L}^3$ & 10.000026 & 0.000030 \\
 $\bar{L}^4$ & 4.9999978 & 2.5\times10^{-6} \\
 $m^2$ & 464.00001 & 1.9\times10^{-6} \\
 $\bar{L} m^2$ & -96.000001 & 3.1\times10^{-6} \\
 $\bar{L}^2 m^2$ & -96.000001 & 9.0\times10^{-7} \\
\hline
\hline
 $1$ & -2.883\times10^{-6} & 8.8\times10^{-7}  \\
 $m$ & -15.999999 & 2.0\times10^{-7} \\
 $\bar{L} m$ & -32.000000 & 7.9\times10^{-8}
\end{tabular}
\caption{Angular fit results from fitting $64{\rm Re}[C_3]$ (upper rows) and ${\rm Im}[C_3]$ (lower rows) to the $n=1$ family of TTMs.  See the caption for Table~\ref{table:C1} for additional details.}
\label{table:C3}
\end{table}

\begin{table}
 \begin{tabular}{c| d{11} d{7}}
 &\multicolumn{1}{c}{Estimate} & \multicolumn{1}{c}{$\sigma$}  \\
\hline
 $1$ & -2.8687\times10^{-6} & 1.6\times10^{-6} \\
 $m$ & 26.000003 & 9.6\times10^{-7} \\
 $\bar{L} m$ & -84.000000 & 4.5\times10^{-7} \\
 $\bar{L}^2 m$ & -84.000000 & 9.5\times10^{-8} \\
 $m^3$ & 7.9999999 & 1.7\times10^{-8} \\
\hline
\hline
 $1$ & -52933.002 & 0.0014 \\
 $\bar{L}$ & -96808.996 & 0.0055 \\
 $\bar{L}^2$ & 27137.995 & 0.0067 \\
 $\bar{L}^3$ & 17982.002 & 0.0030 \\
 $\bar{L}^4$ & -165.00028 & 0.00058 \\
 $\bar{L}^5$ & -65.999983 & 0.000039 \\
 $m^2$ & -29088.000 & 0.00013 \\
 $\bar{L} m^2$ & -56992.000 & 0.00021 \\
 $\bar{L}^2 m^2$ & 3552.0001 & 0.000081 \\
 $\bar{L}^3 m^2$ & 2368.0000 & 0.000011 \\
 $m^4$ & -128.00001 & 2.1\times10^{-6} \\
 $\bar{L} m^4$ & -256.00002 & 2.6\times10^{-6}
\end{tabular}
\caption{Angular fit results from fitting ${\rm Re}[C_4]$ (upper rows) and $1024{\rm Im}[C_4]$ (lower rows) to the $n=1$ family of TTMs.  See the caption for Table~\ref{table:C1} for additional details.}
\label{table:C4}
\end{table}

\begin{widetext}
\begin{align}\label{eqn:asymptotic normal solution}
  \scAnorm{s}{\ell m}{c} = i c (2 \bar{L}+1)
& -\frac14 \left[2 \bar{L} (\bar{L}+1) -4 m^2 -4 s(s-1)+3\right] \nonumber \\
\mbox{} &+\frac{i}{16 c}\left[(2 \bar{L}+1) \left(\bar{L} (\bar{L}+1)-8 m^2-16 s^2+3\right)-32i m s^2 \right] \nonumber \\
\mbox{} & +\frac1{64 c^2}\biggl[5
\left(\bar{L}(\bar{L}+1)(\bar{L}(\bar{L}+1)+7)+3\right)-48 \left(2 \bar{L} (\bar{L}+1)+1\right) m^2 \nonumber\\
\mbox{} &\hspace{70pt}-32 \left(6 \bar{L} (\bar{L}+1)-4 m^2-2 s^2+3\right)s^2-256 i (2 \bar{L}+1)m s^2 \biggr] \nonumber \\
\mbox{} &-\frac{i}{256 c^3}\biggl[
\biggl(\frac14(2\bar{L}+1)\left(\bar{L}(\bar{L}+1)(33\bar{L}(\bar{L}+1)+415)+453\right) \nonumber\\
\mbox{} &\hspace{100pt}-8 (2 \bar{L}+1) \left(37 \bar{L}(\bar{L}+1)+51\right) \left(m^2+2 s^2\right) \nonumber \\
\mbox{} &\hspace{160pt}+32 (2 \bar{L}+1) \left(m^4+60 m^2 s^2+32 s^4\right)\biggr) \nonumber \\
\mbox{} &\hspace{130pt}-256i\left(21 \bar{L} (\bar{L}+1)-2 m^2-4 s^2+\frac{19}2\right)m s^2\biggr]
+ \mathcal{O}(c^{-4})
\end{align}
\end{widetext}

The coefficients $B_3$, $B_4$, and $B_5$ can each be determined once the form for the $C_{p-1}$ coefficient is fixed.  At each order, we find that the coefficients for the two families simply differ by a phase rotation of some integer multiple of $\pi/3$.  Because of this, it is convenient to change notation and express the asymptotic expansion of the two families as:
\begin{widetext}
\begin{subequations}\label{eqn:Omega all}
\begin{align}
\label{eqn:Omega n1}
M\omega_{\ell m1}(a) &= \sum_{p=0}^\infty{\frac{\Omega_p}{a^{(4-p)/3}}}, \\
\intertext{and}
\label{eqn:Omega n2}
M\omega_{\ell m2}(a) &= \sum_{p=0}^\infty{\frac{(-1)^{(1-p)/3}\Omega_p}{a^{(4-p)/3}}}, \\
\intertext{where}
\label{eqn:Omega 0}
\Omega_0 &= -i2^{2/3} 3^{1/3}, \\
\label{eqn:Omega 1}
\Omega_1 &= \frac23(2m - i (2 \bar{L}+1)), \\
\label{eqn:Omega 2}
\Omega_2 &= \frac{8 (2 \bar{L}+1) m - i \left(2\bar{L} (\bar{L}+1)+4 m^2-43\right)}{18\times 2^{2/3} 3^{1/3}}, \\
\label{eqn:Omega 3}
\Omega_3 &= -\frac{40 \left(6 \bar{L} (\bar{L}+1)+4 m^2-3\right)m 
            - i (2 \bar{L}+1) \left(19 \bar{L} (\bar{L}+1)-264 m^2+2149\right)}{1296\times 2^{1/3} 3^{2/3}}, \\
\label{eqn:Omega 4}
\Omega_4 &= \frac{1}{144} (2 \bar{L}+1) \left(\bar{L} (\bar{L}+1)-8 m^2+3\right)m
            - i\frac{3 \bar{L} (\bar{L}+1)(\bar{L} (\bar{L}+1)-13)-32 m^2 \left(m^2-4\right)-69}{1152} ,\\
\label{eqn:Omega 5}
\Omega_5 & -\frac{1}{4\,478\,976\times 2^{2/3} 3^{1/3}}\biggl[ 
32\Bigl(\bar{L}(\bar{L}+1)\bigl(775\bar{L}(\bar{L}+1)+9760m^2-64\,759\bigr) \nonumber \\
& \hspace{220pt} - 5824m^4+36\,688m^2-35\,231\Bigr)m \nonumber \\
& \hspace{90pt} - i(2\bar{L}+1)\Bigl(\bar{L}(\bar{L}+1)\bigl(2915\bar{L}(\bar{L}+1)+42\,208m^2-134\,963\bigr) \nonumber \\
& \hspace{220pt} + 479\,104m^4-1\,433\,696m^2-4\,864\,417\Bigr)
\Biggr].
\end{align}
\end{subequations}
\end{widetext}
Notice that the ${\rm Re}[\Omega_p]$ only contains odd powers of $m$, while in the ${\rm Im}[\Omega_p]$ only even powers of $m$ occur.  This behavior is necessary for the $n=1$ family of modes to be degenerate with their mirror modes.  Furthermore, the ${\rm Re}[\Omega_p]$ is proportional to $m$ so that, in the asymptotic regime, $\omega_{\ell01_0}$ must be purely imaginary. The extra factor of $(-1)^{(1-p)/3}$ in Eq.~(\ref{eqn:Omega n2}), which multiplies each $\Omega_p$ by a phase rotation of some integer multiple of $\pi/3$, breaks this degeneracy for the $n=2$ family of TTMs.  Clearly, from Eqs.~(\ref{eqn:Omega n1}) and (\ref{eqn:Omega n2}), we see that any sequences of solutions obeying these asymptotic behaviors should be considered as belonging, respectively, to either the $n=1$ or $n=2$ families.  This justifies our relabeling of the overtone multiplets $\{\ell,0,0_2\}$ which were treated as a new branch of the $\{\ell,0,0\}$ sequences in Ref.~\cite{cook-et-al-2018}, as the $\{\ell,0,1_0\}$ sequences in the new $n=1$ family of TTMs.

\begin{figure}[h]
\begin{tabular}{c}
    \includegraphics[width=\linewidth,clip]{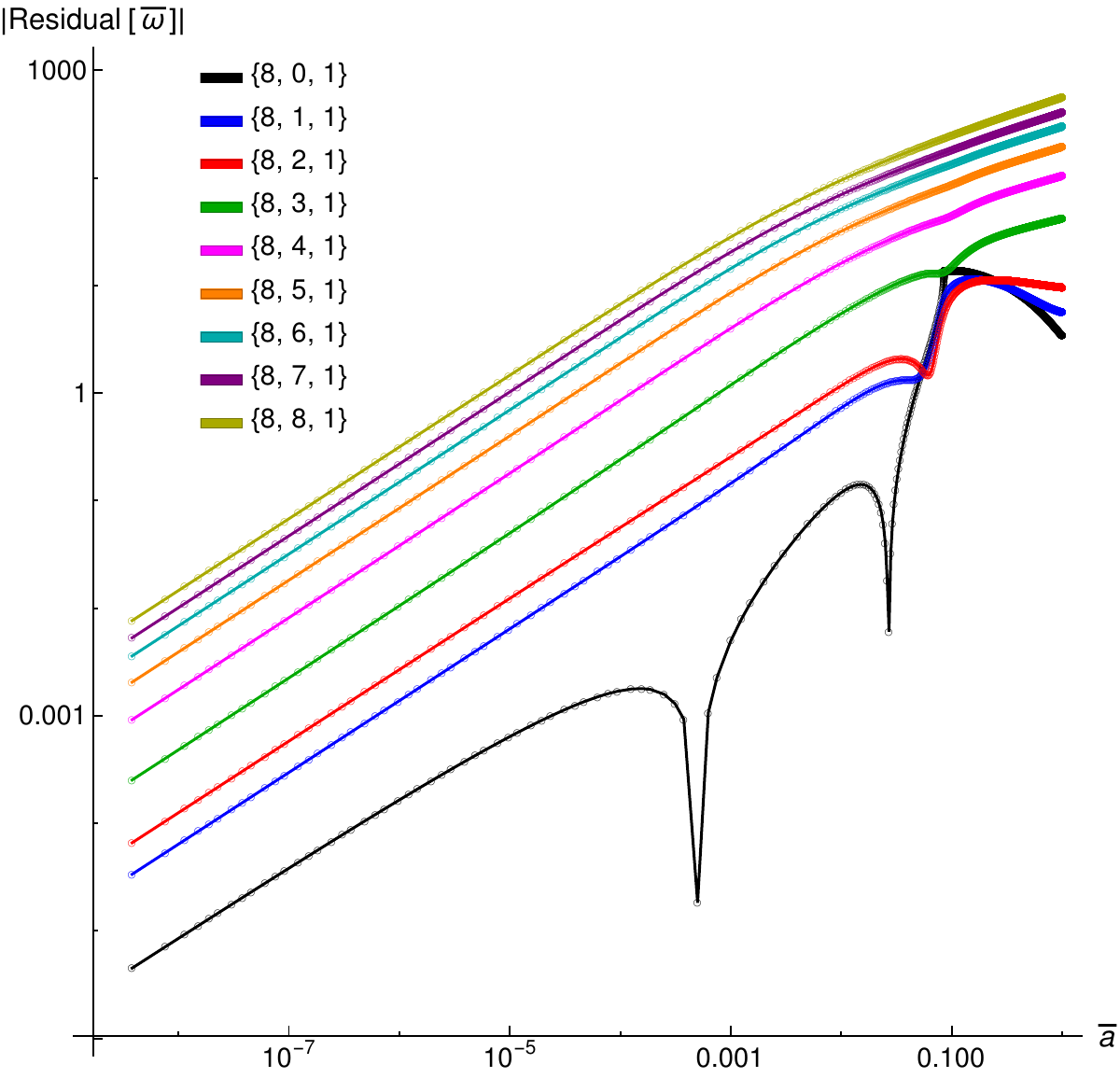} \\
    \includegraphics[width=\linewidth,clip]{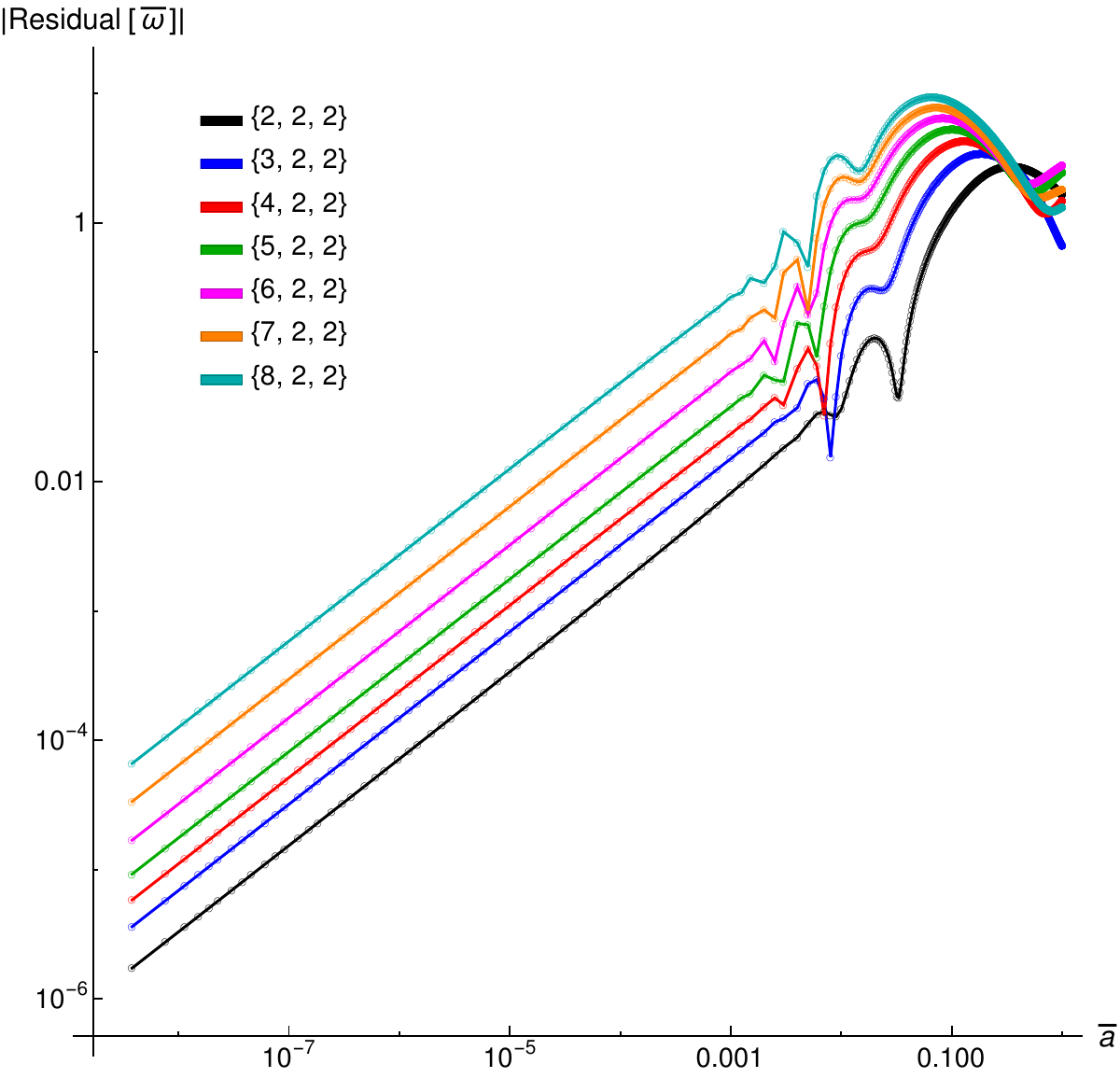}
\end{tabular}
    \caption{Log-log plots of the residual of $\bar\omega$.  The residual is defined as the difference of Eq.~(\ref{eqn:Omega n1}) or (\ref{eqn:Omega n2}) with the corresponding data from the numerical solutions for the TTMs.  The upper panel shows a selection of residuals for the $n=1$ family of TTMs, while the lower panel shows a selection from the $n=2$ family.  In both cases, the asymptotic portion of the the sequences should have a slope of $2/3$, and these representative plots show good agreement with this.}  
    \label{fig:residuals}
\end{figure}

The fidelity of the two asymptotic expansions given in Eq.~(\ref{eqn:Omega all}) can best be illustrated by examining the residuals of the asymptotic expansions.  We define the residual as the magnitude of the difference between the appropriate asymptotic expansion from Eq.~(\ref{eqn:Omega all}) and the corresponding numerical data.  In Fig.~\ref{fig:residuals}, we plot the magnitude of the residual as a function of $\bar{a}$ on a log-log plot.
Because the first unknown coefficient in the asymptotic expansions is at order $\bar{a}^{2/3}$, we should expect the slope of the magnitude of the residual in the asymptotic regime to be approximately $2/3$.  This is, in fact what we find in all cases.  Measuring the slope of the asymptotic region using the first $4$ data points in each sequence, we find that most of the slopes deviate from $2/3$ by less than $0.1\%$.  The worst case occurs for the $\{7,0,1_0\}$ sequence which deviates by $2.6\%$.  The $\{\ell,0,1_0\}$ sequences are all more susceptible to larger deviations in the measured slope because ${\rm Re}[\bar\omega]=0$, making it easy for the residual to actually pass through zero.  In the upper panel of Fig.~\ref{fig:residuals}, we see the $\{8,0,1_0\}$ sequence displays a zero crossing at $\bar{a}$ just less than $0.001$, causing it to have a measured asymptotic slope deviating from $2/3$ by $1.3\%$.  The $\{7,0,1_0\}$ sequence has a similar zero crossing just below $\bar{a}=10^{-4}$.  The $16$ example sequence residuals shown in Fig.~\ref{fig:residuals} are very representative of the behavior of all sequence residuals and clearly show that the asymptotic behavior of the two new families of TTMs is correctly modeled by the expressions in Eq.~(\ref{eqn:Omega all}).  Furthermore, since the asymptotic expansions of $\bar\omega$ are based directly on the asymptotic expansion of the separation constant given in Eq.~(\ref{eqn:asymptotic normal solution}), it follows that this should show analogous behavior of its residuals.  We have examined these, but do not present any of these figures here.

\subsection{Asymptotic behavior of $\scA{s}{\ell m}{c}$}
\label{sec:behavior of A}
To our knowledge, prior to this work no general expression for the asymptotic behavior of $\scA{s}{\ell m}{c}$ for complex values of $c$ has appeared in the literature.  While the success of Eq.~(\ref{eqn:asymptotic normal solution}) is encouraging, some words of caution are warranted.  

First, we remind the reader that Eq.~(\ref{eqn:asymptotic normal solution}) is identical to Eq.~(26) from Ref.~\cite{VickersCook2022} which gives the asymptotic behavior for purely imaginary values of $c$.  However, apart from the $s$ behavior, the form of Eq.~(\ref{eqn:asymptotic normal solution}) has been fit independently using data with complex values of $c$.  While we cannot completely rule out that the general asymptotic expansion has different coefficients multiplying even powers of $s$ than those we present in Eq.~(\ref{eqn:asymptotic normal solution}), we find this possibility very unlikely since we expect the expansions to agree in the limit that $c$ approaches purely imaginary values.

\begin{figure}[h]
\begin{tabular}{cc}
    \includegraphics[width=0.5\linewidth,clip]{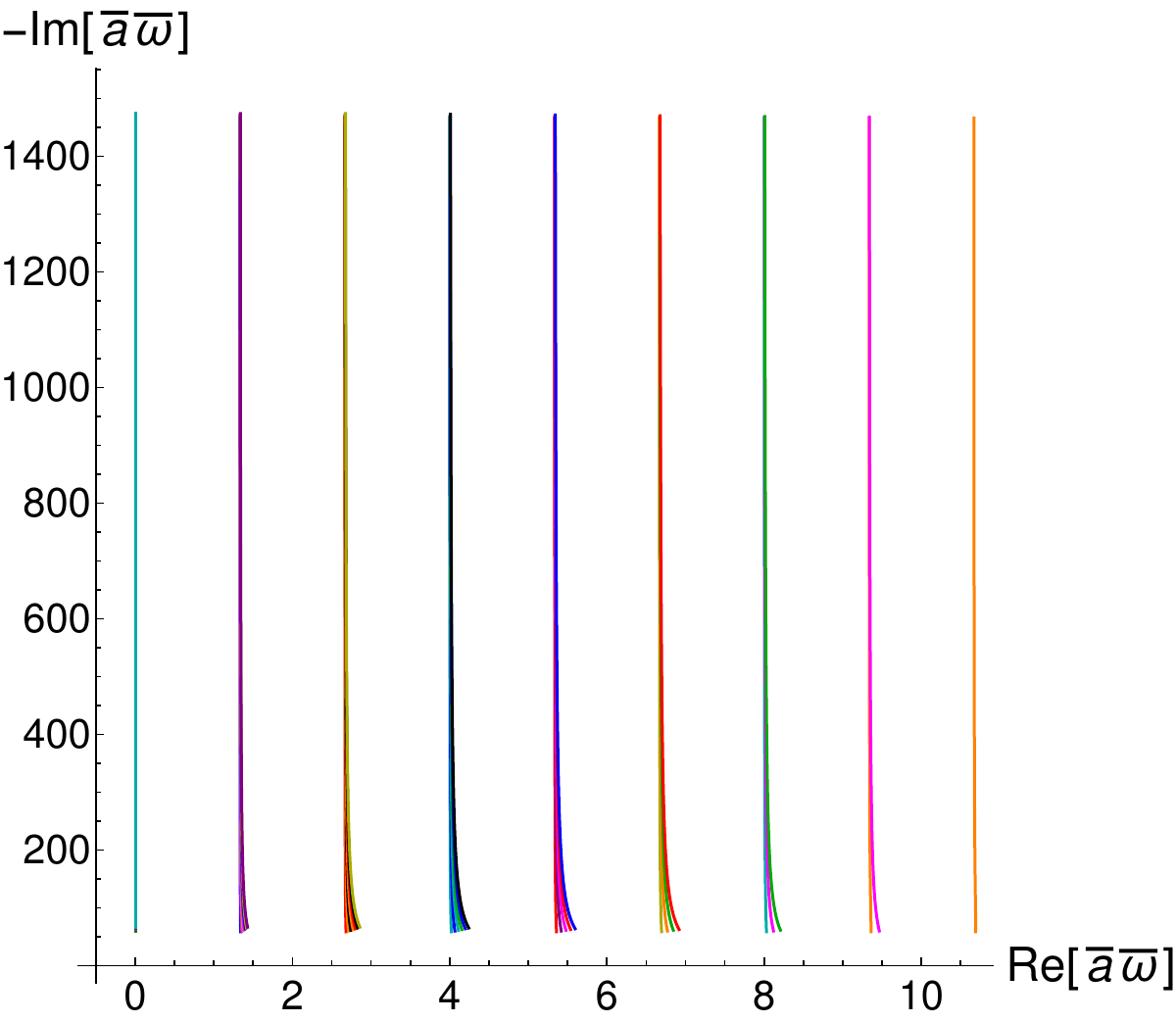} &
    \includegraphics[width=0.5\linewidth,clip]{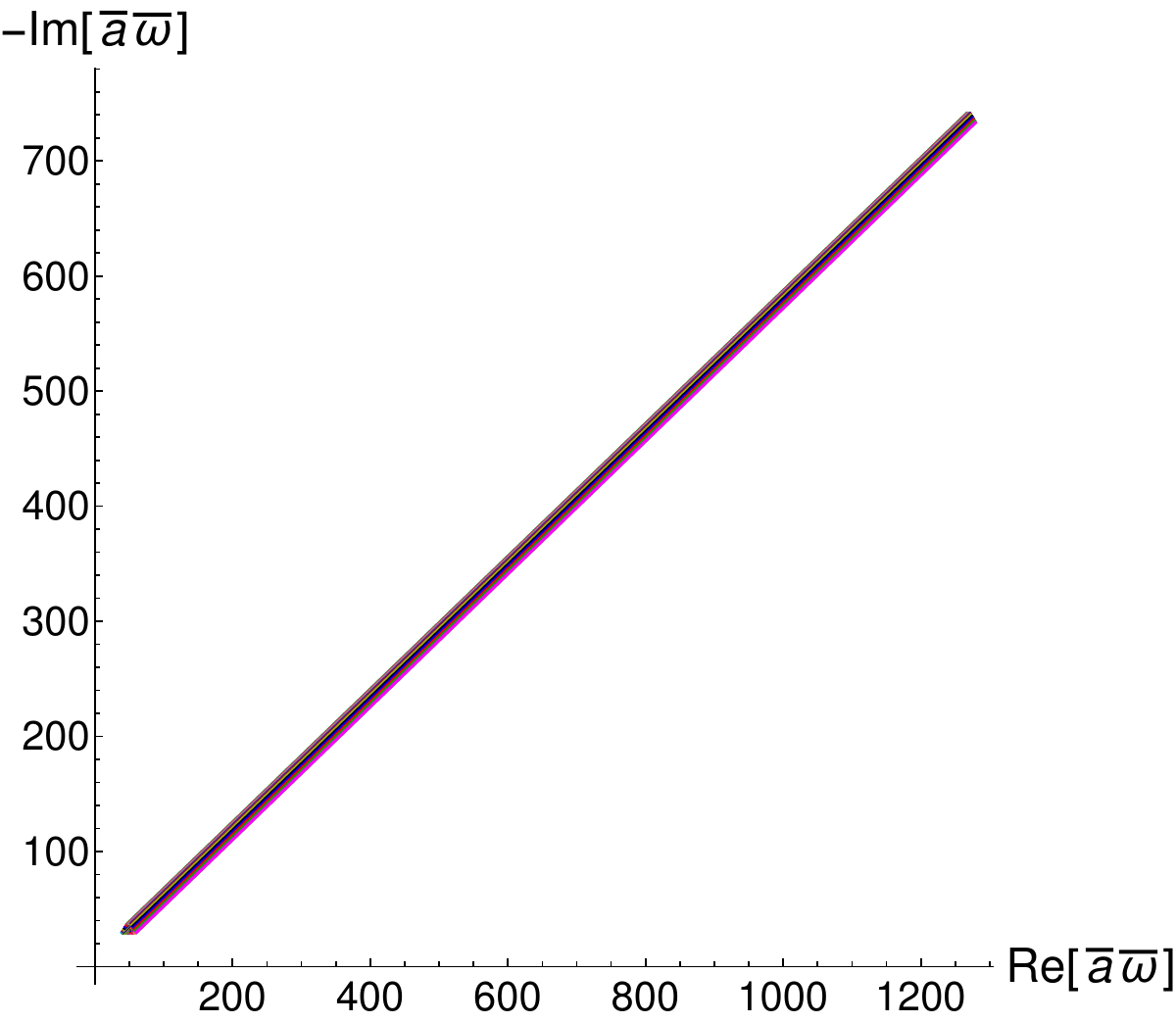}
\end{tabular}
    \caption{The two panels display the range of complex values of $c=\bar{a}\bar\omega$ sampled by the two new families of TTMs.  The left panel corresponds to the $n=1$ family, while the right panel corresponds to the $n=2$ family.}
    \label{fig:c values}
\end{figure}

A second caveat is that our TTM sequences only cover a limited range of possible complex values of $c$.  Figure~\ref{fig:c values} displays all of the values of $c=\bar{a}\bar\omega$ covered by all of our new TTM sequences used to determine the asymptotic expansion for $\scA{s}{\ell m}{c}$.
The left panel of Fig.~\ref{fig:c values} corresponds to the $n=1$ family of TTMs for which we see that, asymptotically, ${\rm Re}[\bar{a}\bar\omega]=\frac43m$.  The right panel corresponds to the $n=2$ family of TTMS.  In this case, all of the sequences have an asymptotic slope for $\bar{a}\bar\omega$ of $1/\sqrt{3}$.  The intercepts of the asymptotic behavior differ, so these sequences cover a narrow path in the complex $c$ plane.  Continuity of the solution again suggests that Eq.~(\ref{eqn:asymptotic normal solution}) should be correct for all asymptotic complex values of $c$ with ${\rm Re}[c]>0$ and ${\rm Im}[c]<0$.  The restriction to a particular quadrant of the complex $c$ plane is because Eq.~(\ref{eqn:asymptotic normal solution}) does not satisfy Eqs.~(\ref{eqn:swSF_mcA_ident}) and (\ref{eqn:swSF_cA_ident}).  These identities must be explicitly imposed in order to evaluate the asymptotic behavior in the other three quadrants of the complex $c$ plane.

The final, and most important, caveat is that the asymptotic behavior for $\scA{s}{\ell m}{c}$ seen in the two new families of TTMs may not be the only possible asymptotic behavior.  In Ref.~\cite{VickersCook2022}, it was discovered that two possible asymptotic behaviors exist for purely imaginary values of $c$ (the prolate case).  In this prolate case, most sequences of $\scA{s}{\ell m}{c}$ behave asymptotically as described by Eq.~(\ref{eqn:asymptotic normal solution}), and such sequences are referred to as ``normal.''  However, some sequences display a distinctly different leading order asymptotic behavior and are referred to as ``anomalous.''  This anomalous behavior in the prolate asymptotic limit is very similar to the oblate (purely real $c$) asymptotic behavior of $\scA{s}{\ell m}{c}$.  It seems very likely that anomalouslike behavior will also be found along some asymptotic sequences with general complex values of $c$\cite{Barrowes-etal-2004}.

\section{Discussion}
\label{sec:discussion}

In this work, we have shown that the $s=\pm2$ gravitational TTMs of the Kerr geometry contain a much richer structure than was previously suspected.  The original family of mode sequences, labeled by $n=0$ in this paper, exist with values of the mode frequency $\omega$ which are always finite and connect to a Schwarzschild limit with mode frequencies given by Eq.~(\ref{eqn:alg-spec-sch}).  Moving along each sequence as $\bar{a}$ is increased to the extreme limit, we find that the mode frequencies converge toward the vicinity of $\omega=0$.

Both of the new families of TTMs explored in this paper share this latter behavior with their mode frequencies approaching the origin as $\bar{a}\to1$.  On the other hand, the two new families of mode sequences, labeled by $n=1$ and $n=2$, share the distinctly odd behavior that their Schwarzschild limit has mode frequencies that reside at complex infinity.  This behavior is clearly seen in the various figures in this paper and is made quantitative in the asymptotic expansions given by Eqs.~(\ref{eqn:Omega n1}) and (\ref{eqn:Omega n2}).

Interestingly, one of the new families($n=1$) shares with the original family($n=0$) the behavior that its mirror-mode solutions are degenerate.  That is, $\omega^-_{\ell mn}=\omega^+_{\ell mn}$ for $n\in\{0,1\}$.  In contrast, the other new family($n=2$) presents a full set of nondegenerate modes similar to what is seen with the QNMs.  An interesting behavior of the QNMs is that no examples of sequences have been found which cross the NIA.  In cases with $|m|>0$, no QNMs have been found with a purely imaginary frequency.\footnote{See Ref.~\cite{cook-zalutskiy-2016b} for an exploration of the behavior of QNMs at the NIA, and in particular Sec.~IV.B for as discussion of the limitations on QNMs at the NIA.}  But, for the $n=2$ family of TTMs, we find that some of the sequences of mode frequencies with $|m|>0$ do cross the NIA so that, in some cases, the real parts of both the $\omega^\pm$ solution sets can change sign.

An important question to consider relates to the physical importance and properties of these TTMs.  On this, there is much less that can be said with confidence.  There is some work that explores the importance of QNMs and TTMs in the transition between classical and quantum gravity\cite{Hod-1998,KeshetNeitzke2008}.  In standard quantum theory, the frequencies of TTMs entering a system coincide with the system's metastable eigenfrequencies, and it has been suggested that the TTM frequencies of a black hole could coincide with the eigenenergies of some internal black-hole degrees of freedom\cite{KeshetNeitzke2008}.  But, the only firm physical interpretation is that the TTMs represent linear perturbations of the Kerr metric that effectively travel radially through the spacetime without reflection.

A necessary byproduct of constructing the asymptotic expansion for the TTM mode frequencies was the construction of an analytic asymptotic expansion for the separation constant of the angular Teukolsky equation given by Eq.~(\ref{eqn:swSF_DiffEqn}).  Remarkably, the asymptotic expansions we obtained associated with both the $n=1$ and $n=2$ families of TTMs were identical.  This expansion, given in Eq.~(\ref{eqn:asymptotic normal solution}), is only confirmed to be valid for $s=\pm2$ but is likely valid for all values of $s$.  In addition, Eq.~(\ref{eqn:asymptotic normal solution}) must be evaluated for the quadrant of the complex plane for which $\rm{Re}[c]>0$ and $\rm{Im}[c]<0$, but it can be extended to all values of $c$ by means of Eqs.~(\ref{eqn:swSF_mcA_ident}) and (\ref{eqn:swSF_cA_ident}).

As discussed in more detail in Sec.~\ref{sec:behavior of A}, Eq.~(\ref{eqn:asymptotic normal solution}) may not represent the only possible asymptotic behavior for $\scA{s}{\ell m}{c}$.  As shown in Ref.~\cite{VickersCook2022}, it is possible that the asymptotic behavior of the separation constant will be proportional to $c^2$ rather than to $c$ in some cases.  This raises the possibility that yet another family (or set of families) of TTMs may exist.  As discussed in Sec.~\ref{sec:asymptotic behavior}, assuming asymptotic behaviors given by Eqs.~(\ref{eqn:omega fit function}) and (\ref{eqn:A vs c fit}) leads to only two possible families of TTMs which agree with the numerical solutions for the new $n=1$ and $n=2$ families.  By modifying Eq.~(\ref{eqn:A vs c fit}) to include a leading order term proportional to $c^2$, we have attempted to obtain a new family of solutions.  However, it is unclear how Eq.~\ref{eqn:omega fit function}) should be modified to obtain a reasonable solution, and our initial attempts to find additional new TTMs have not been successful.

\acknowledgments 
Some computations were performed using the Wake Forest University (WFU) High Performance Computing Facility, a centrally managed computational resource available to WFU researchers including faculty, staff, students, and collaborators\cite{DEAC-Cluster}.

\appendix
\section{TTMs with $4\le\ell\le8$}
\label{sec:appendix figures}

\begin{widetext}
In this appendix, we present the plots of the mode frequencies for all three families of TTMs for the cases of $4\le\ell\le8$.  The plots for the original $n=0$ families show the mode frequencies over the full range of $\bar{a}$.  For the two new families, the main part of each figure shows the mode frequencies for the large-$\bar{a}$ end of each sequence and an inset figure illustrates the asymptotic behavior of each sequence as $\bar{a}$ approaches the Schwarzschild limit of $\bar{a}=0$.

\begin{figure}[h]
\begin{tabular}{cc}
    \includegraphics[width=0.5\linewidth,clip]{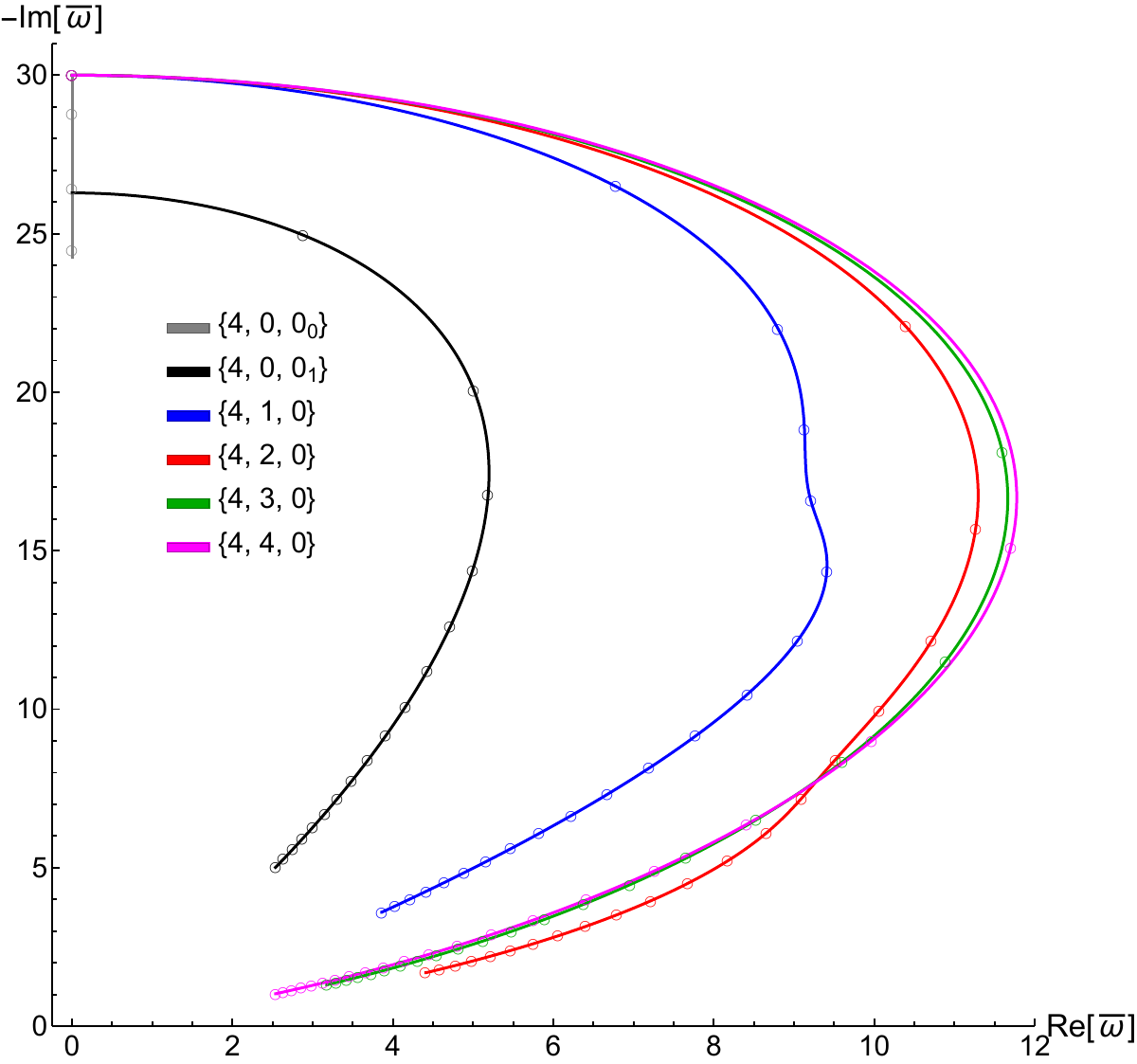} &
    \includegraphics[width=0.5\linewidth,clip]{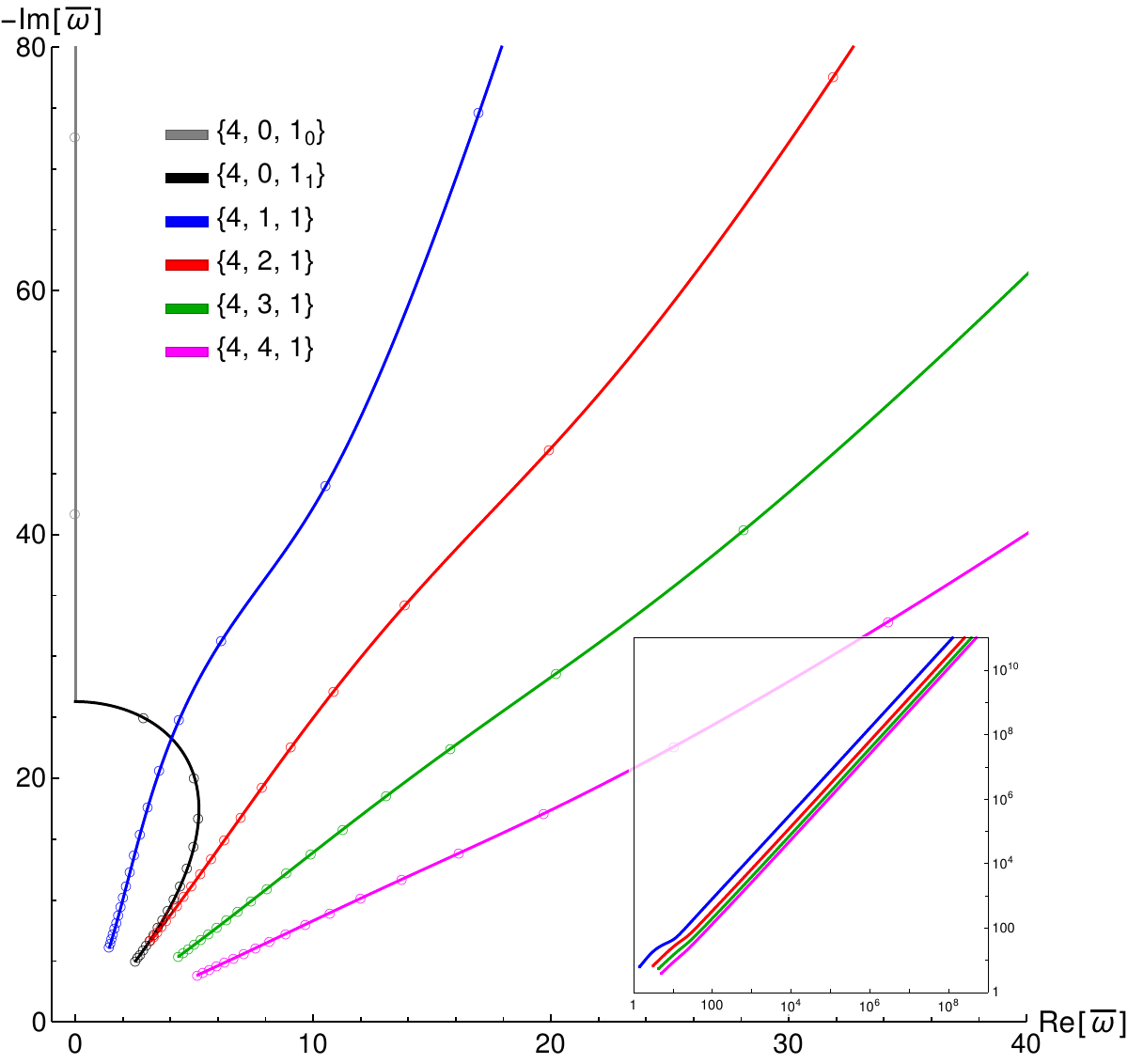} \\
    \multicolumn{2}{c}{\includegraphics[width=0.5\linewidth,clip]{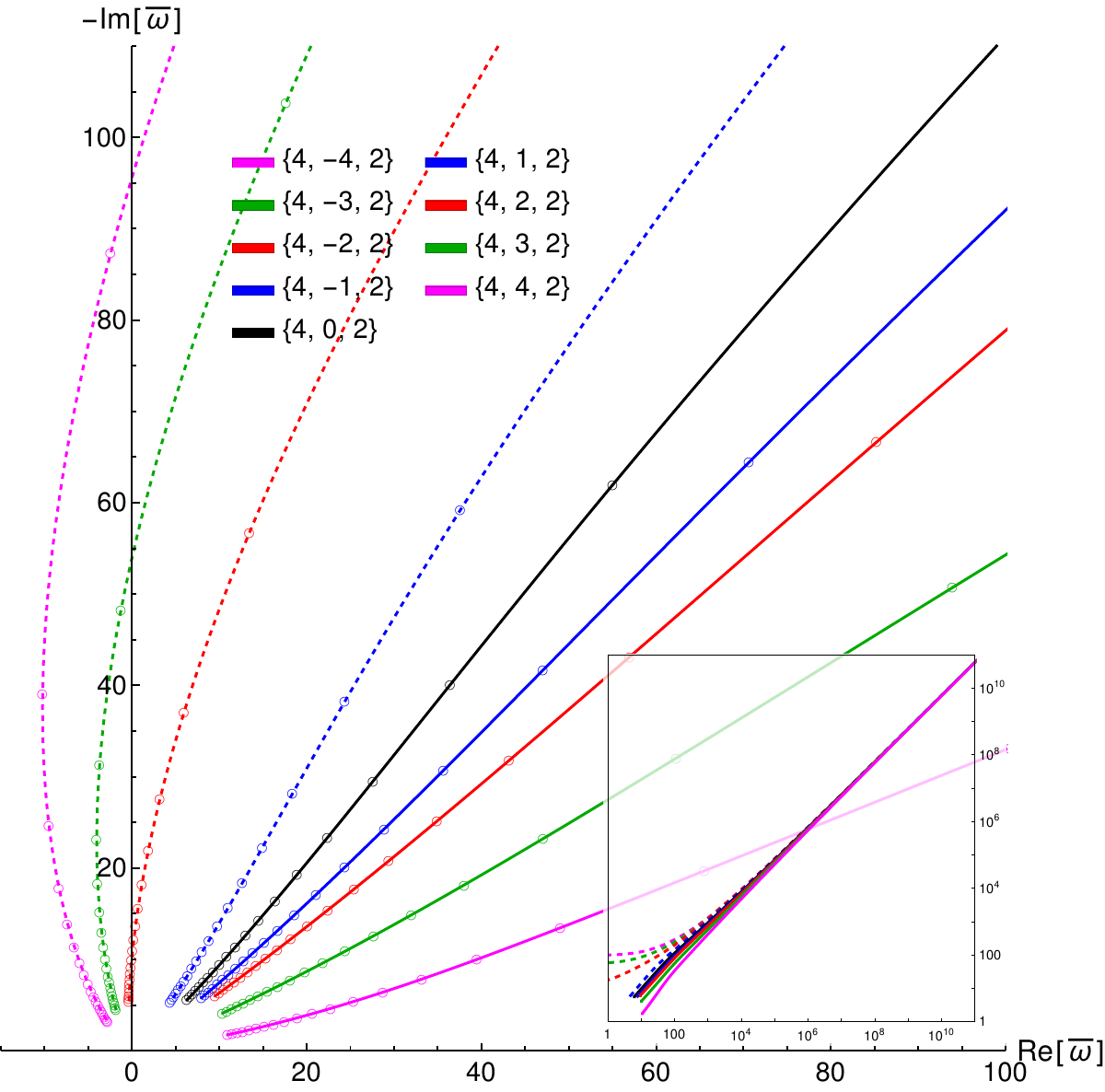}}
\end{tabular}
    \caption{Kerr TTM mode sequences for $\ell=4$. The original family denoted by $n=0$ is in the upper left plot.  The first new family denoted by $n=1$ is in the upper right plot.   The second new family denoted by $n=2$ is in the lower plot.  Mode sequences with negative values of $m$ are drawn as dashed lines.}
    \label{fig:l4 n012 all m}
\end{figure}

\begin{figure}[h]
\begin{tabular}{cc}
    \includegraphics[width=0.5\linewidth,clip]{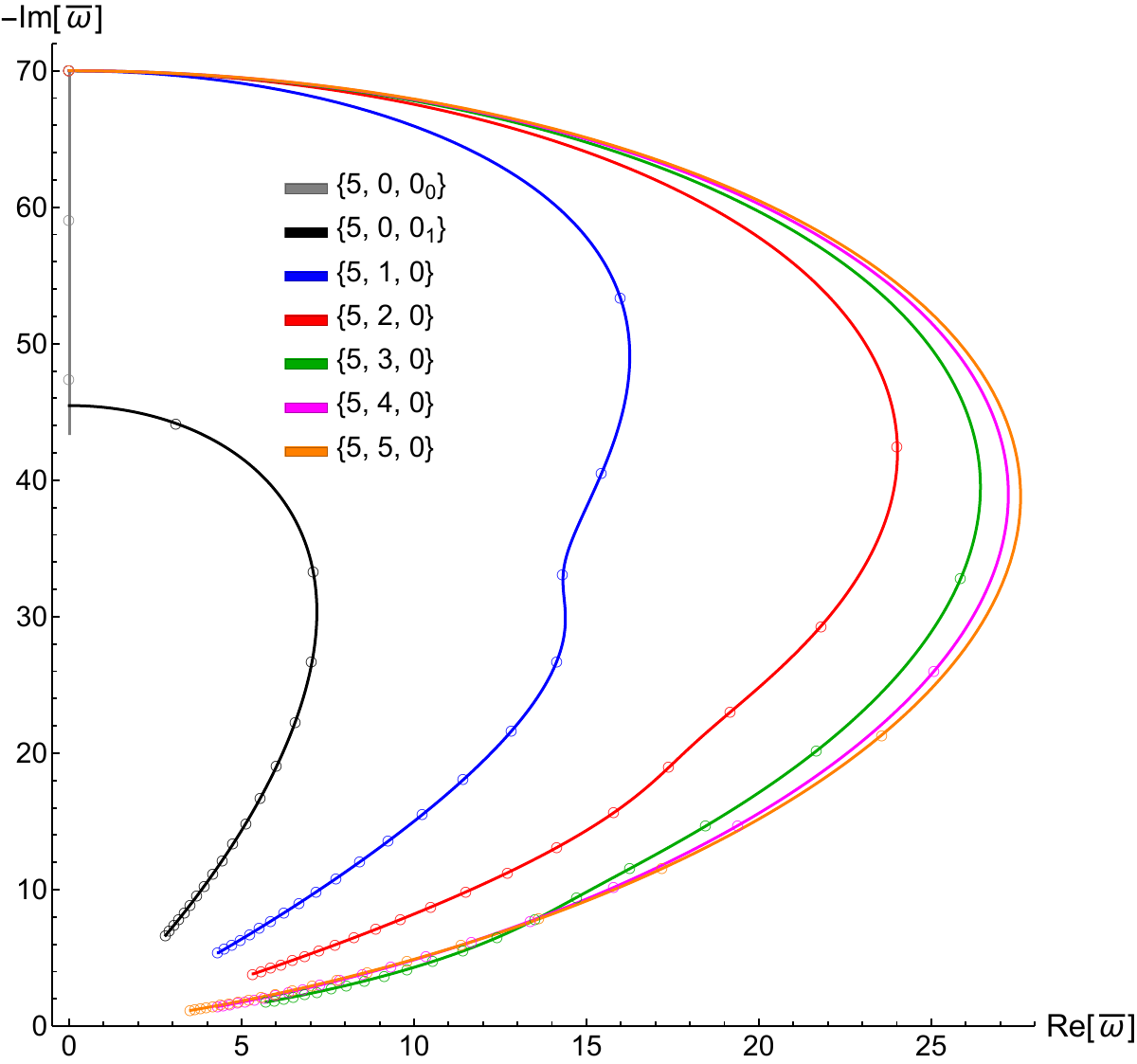} &
    \includegraphics[width=0.5\linewidth,clip]{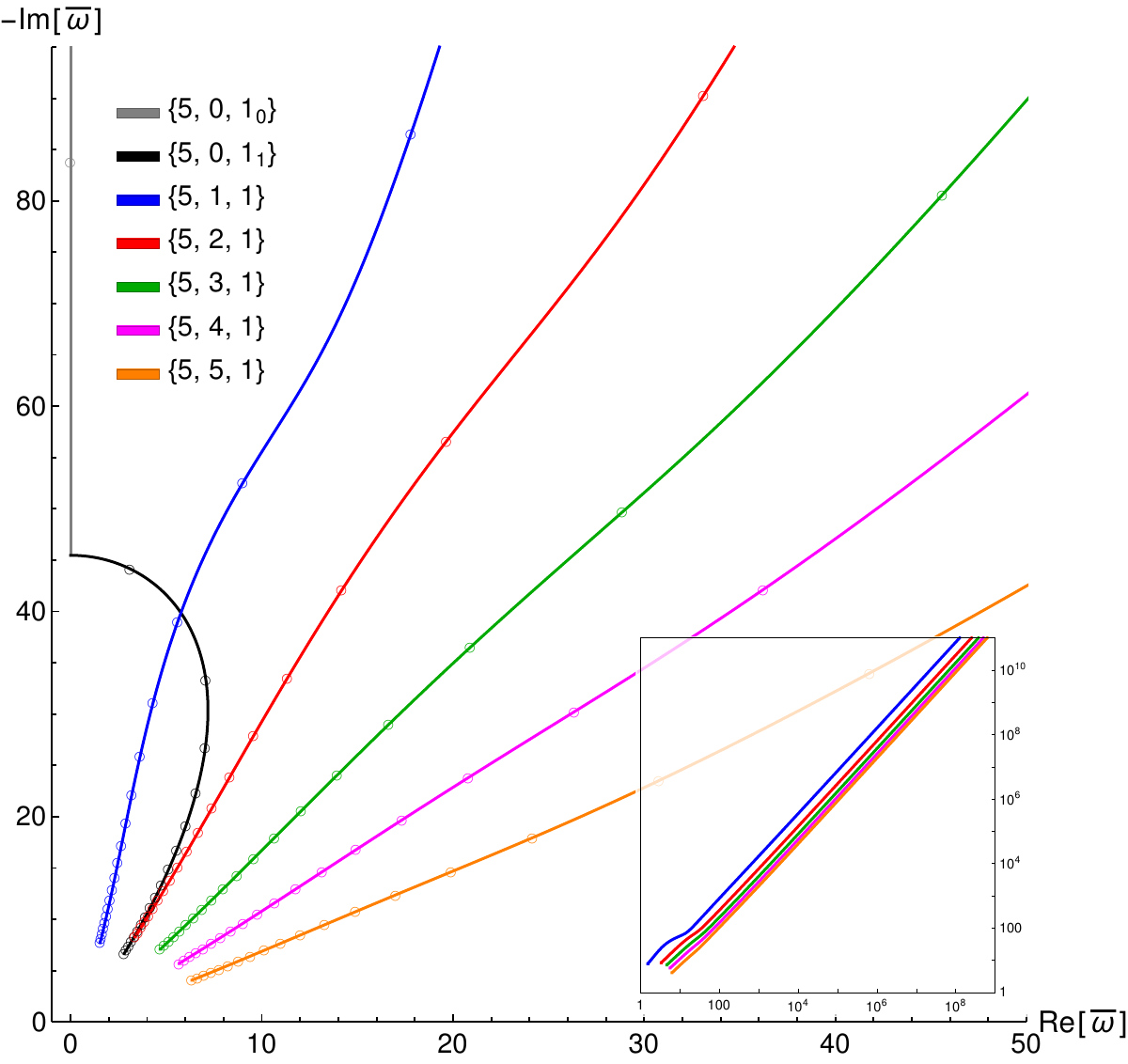} \\
    \multicolumn{2}{c}{\includegraphics[width=0.5\linewidth,clip]{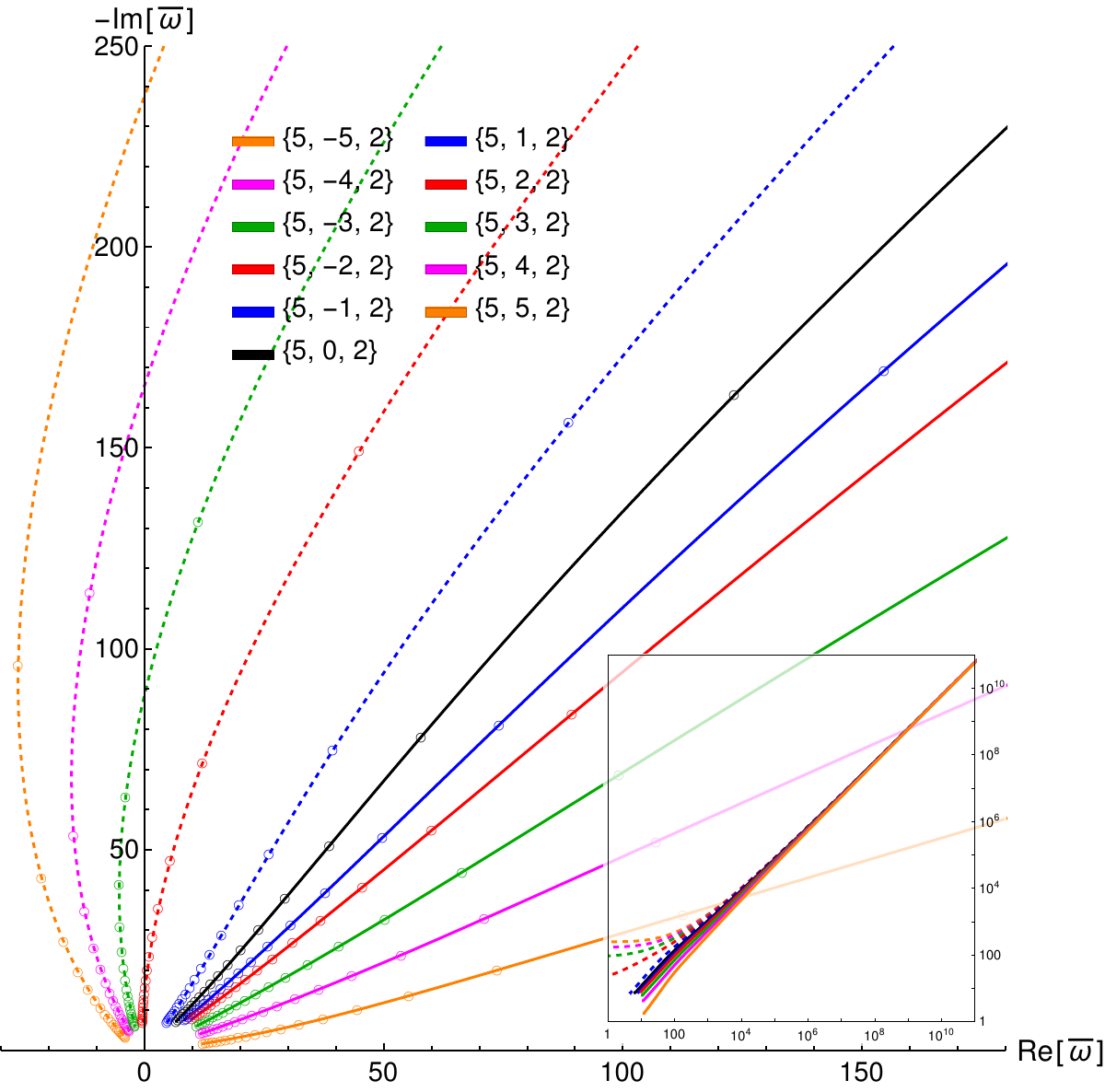}}
\end{tabular}
    \caption{Kerr TTM mode sequences for $\ell=5$. The original family denoted by $n=0$ is in the upper left plot.  The first new family denoted by $n=1$ is in the upper right plot.   The second new family denoted by $n=2$ is in the lower plot.  Mode sequences with negative values of $m$ are drawn as dashed lines.}
    \label{fig:l5 n012 all m}
\end{figure}

\begin{figure}[h]
\begin{tabular}{cc}
    \includegraphics[width=0.5\linewidth,clip]{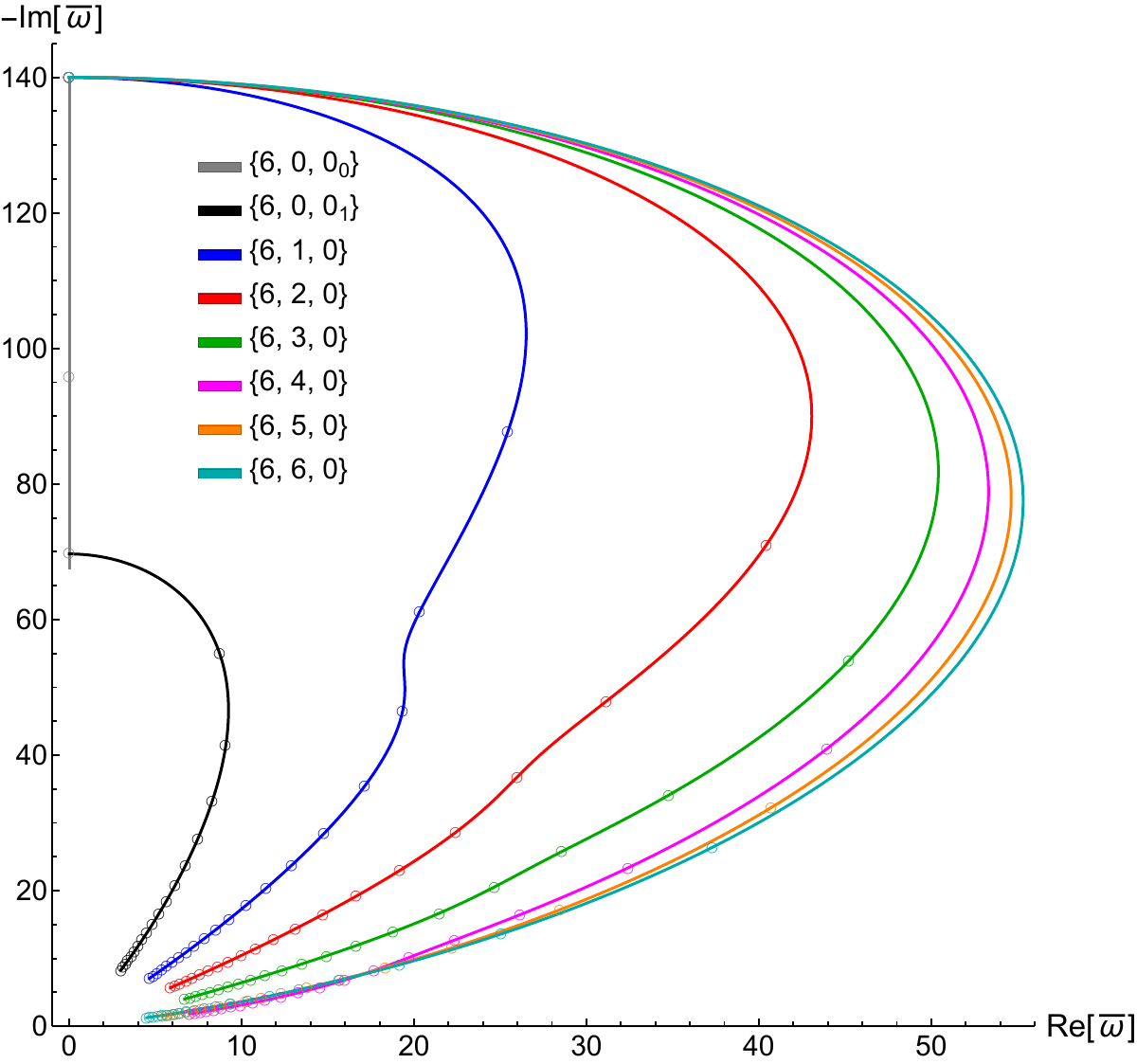} &
    \includegraphics[width=0.5\linewidth,clip]{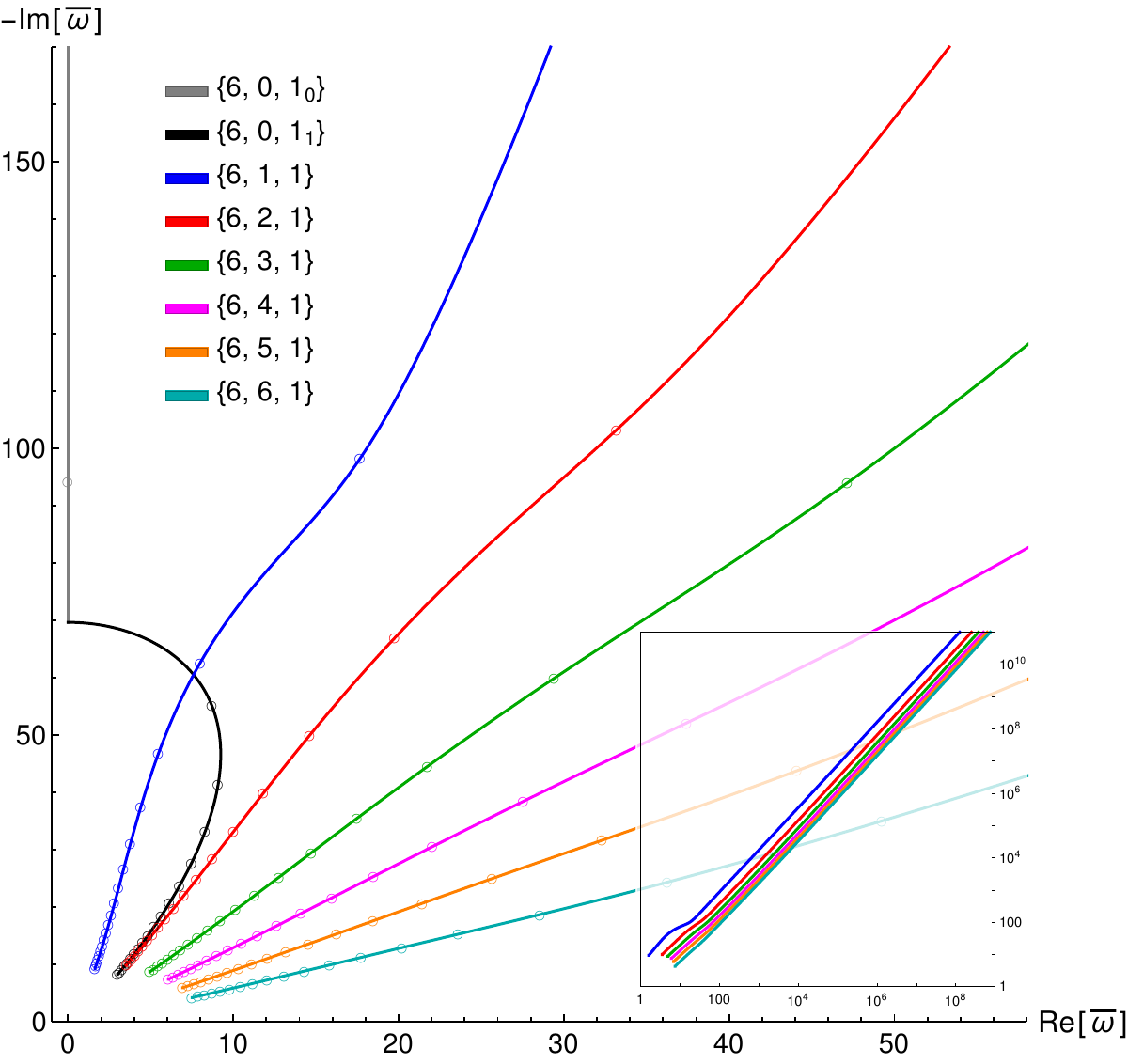} \\
    \multicolumn{2}{c}{\includegraphics[width=0.5\linewidth,clip]{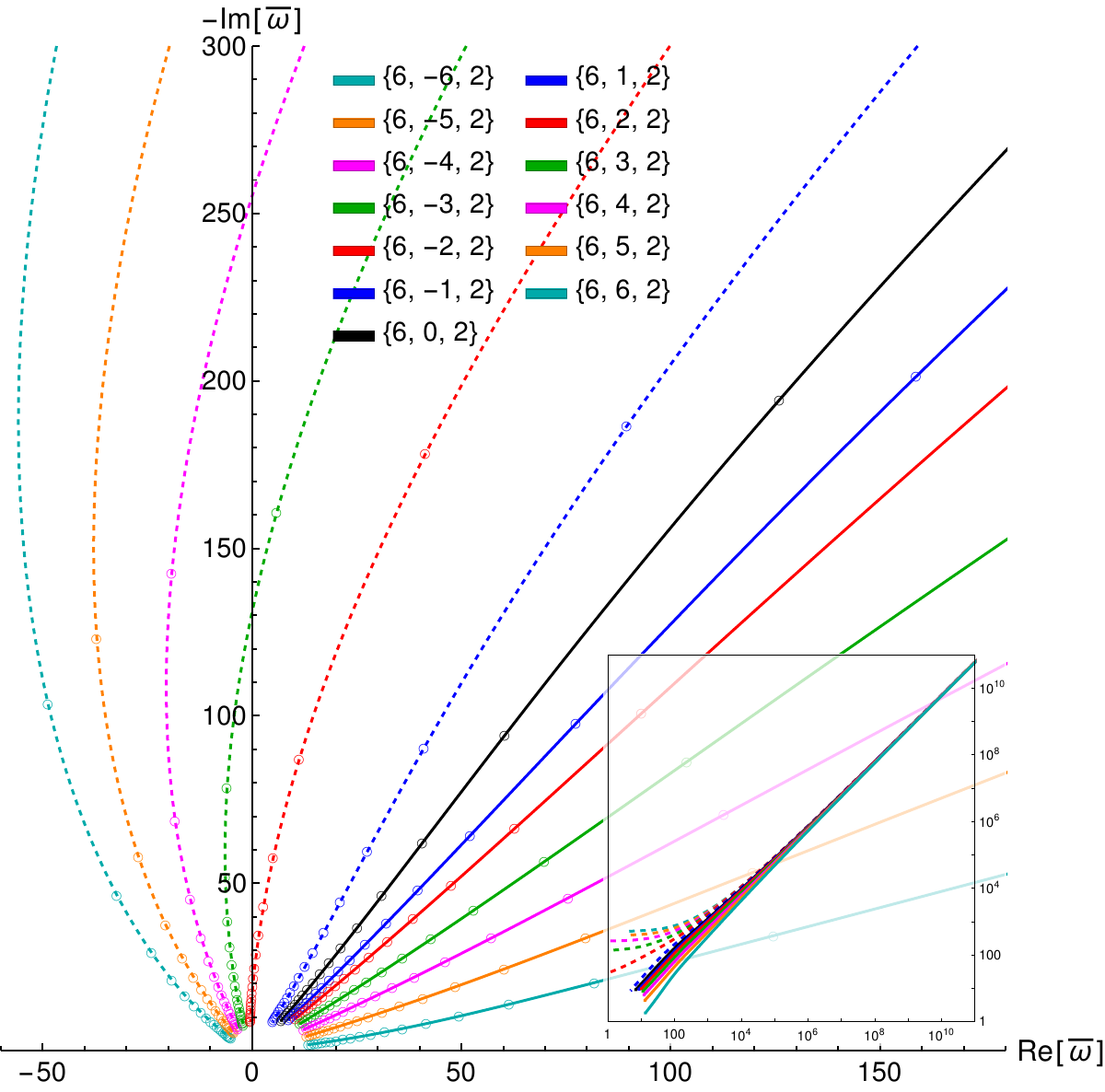}}
\end{tabular}
    \caption{Kerr TTM mode sequences for $\ell=6$. The original family denoted by $n=0$ is in the upper left plot.  The first new family denoted by $n=1$ is in the upper right plot.   The second new family denoted by $n=2$ is in the lower plot.  Mode sequences with negative values of $m$ are drawn as dashed lines.}
    \label{fig:l6 n012 all m}
\end{figure}

\begin{figure}[h]
\begin{tabular}{cc}
    \includegraphics[width=0.5\linewidth,clip]{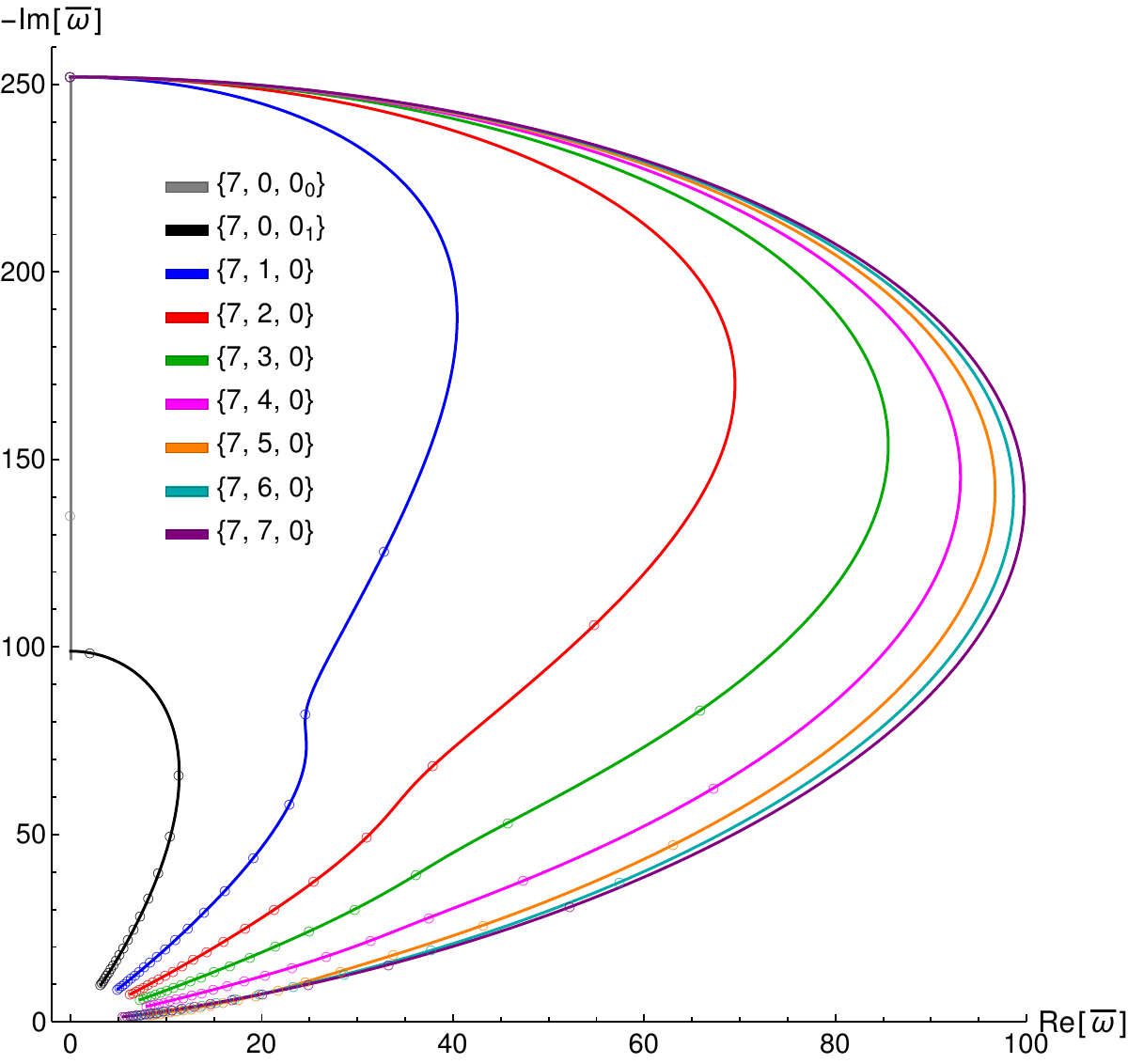} &
    \includegraphics[width=0.5\linewidth,clip]{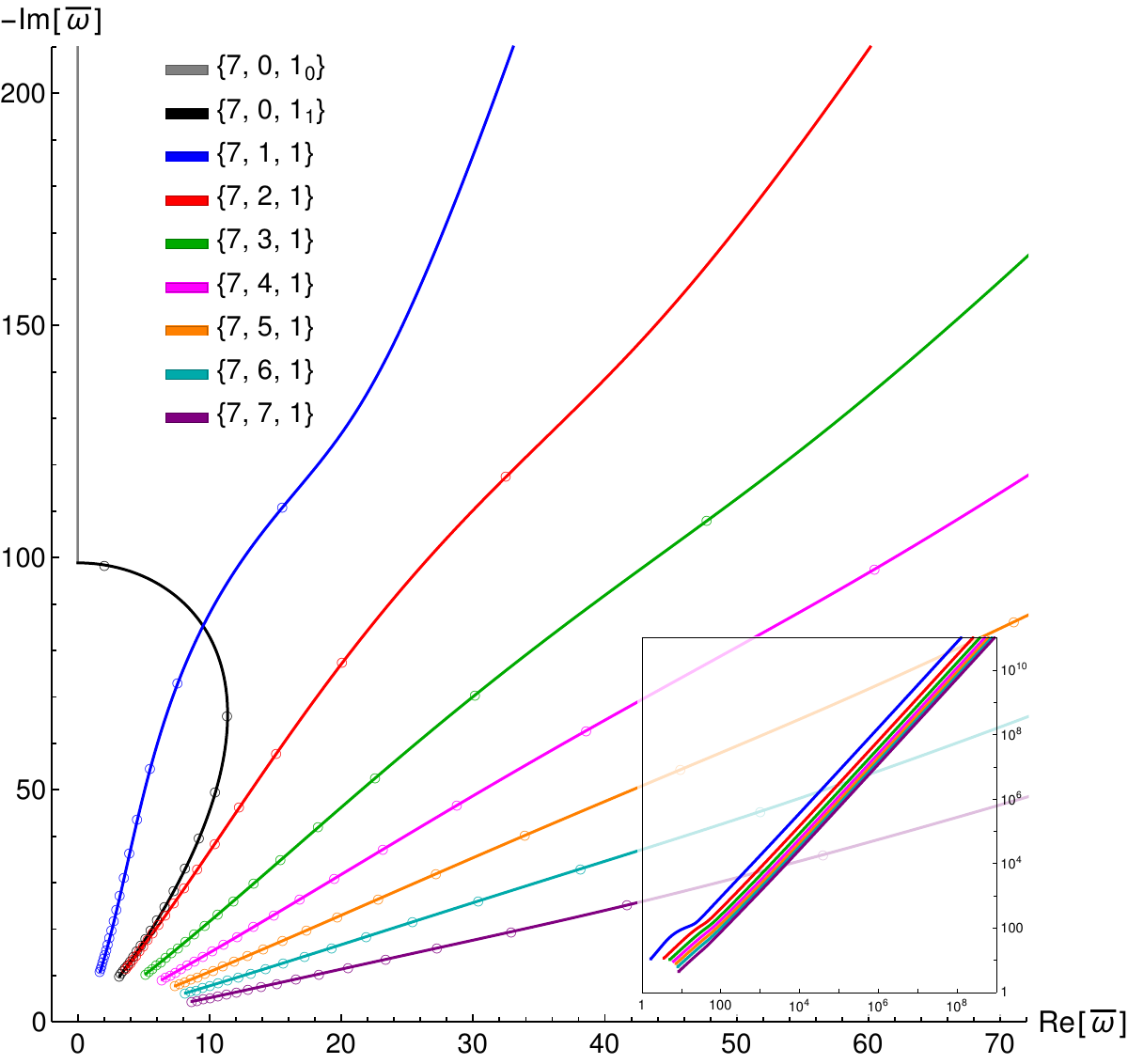} \\
    \multicolumn{2}{c}{\includegraphics[width=0.5\linewidth,clip]{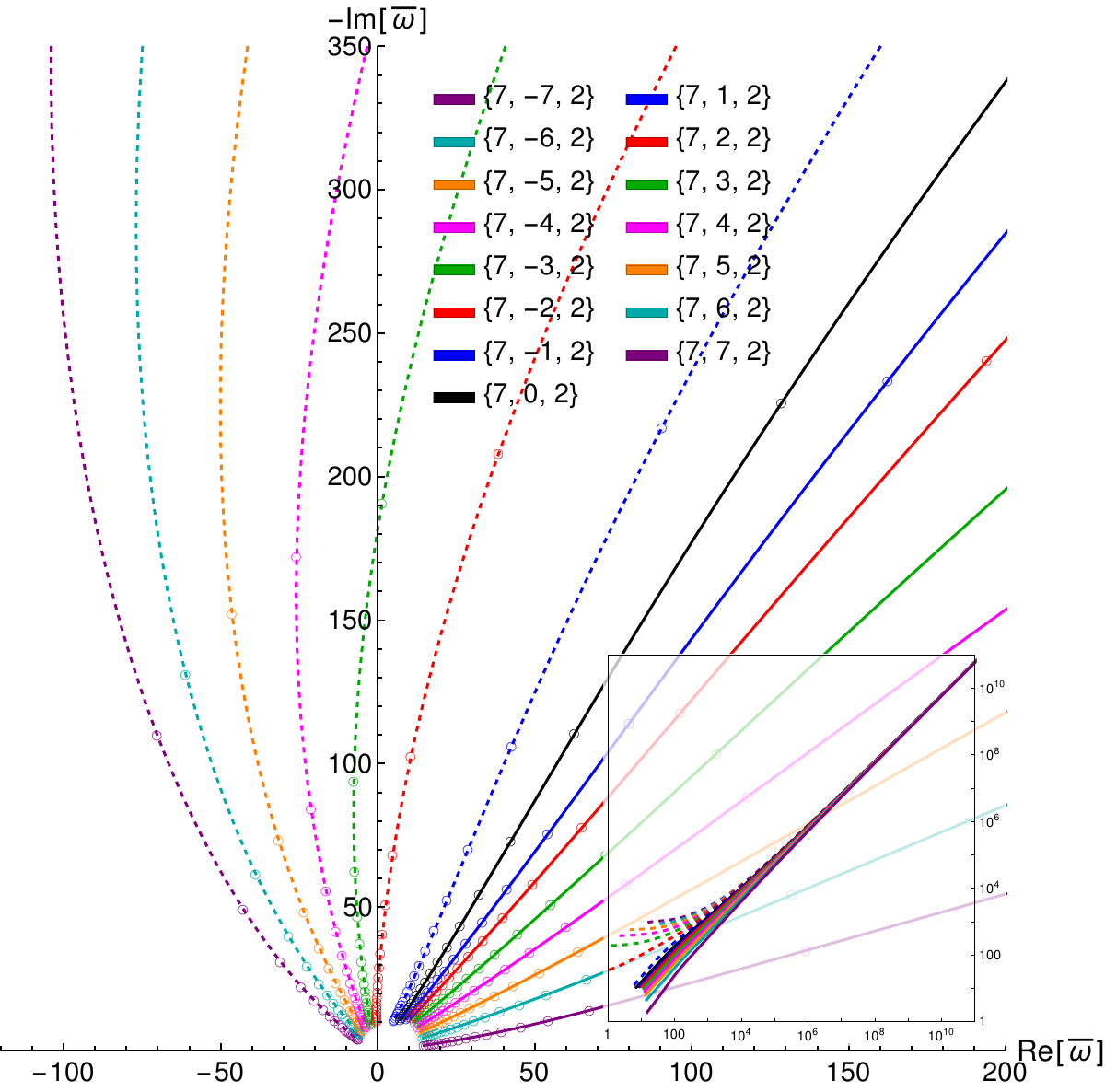}}
\end{tabular}
    \caption{Kerr TTM mode sequences for $\ell=7$. The original family denoted by $n=0$ is in the upper left plot.  The first new family denoted by $n=1$ is in the upper right plot.   The second new family denoted by $n=2$ is in the lower plot.  Mode sequences with negative values of $m$ are drawn as dashed lines.}
    \label{fig:l7 n012 all m}
\end{figure}

\begin{figure}[h]
\begin{tabular}{cc}
    \includegraphics[width=0.5\linewidth,clip]{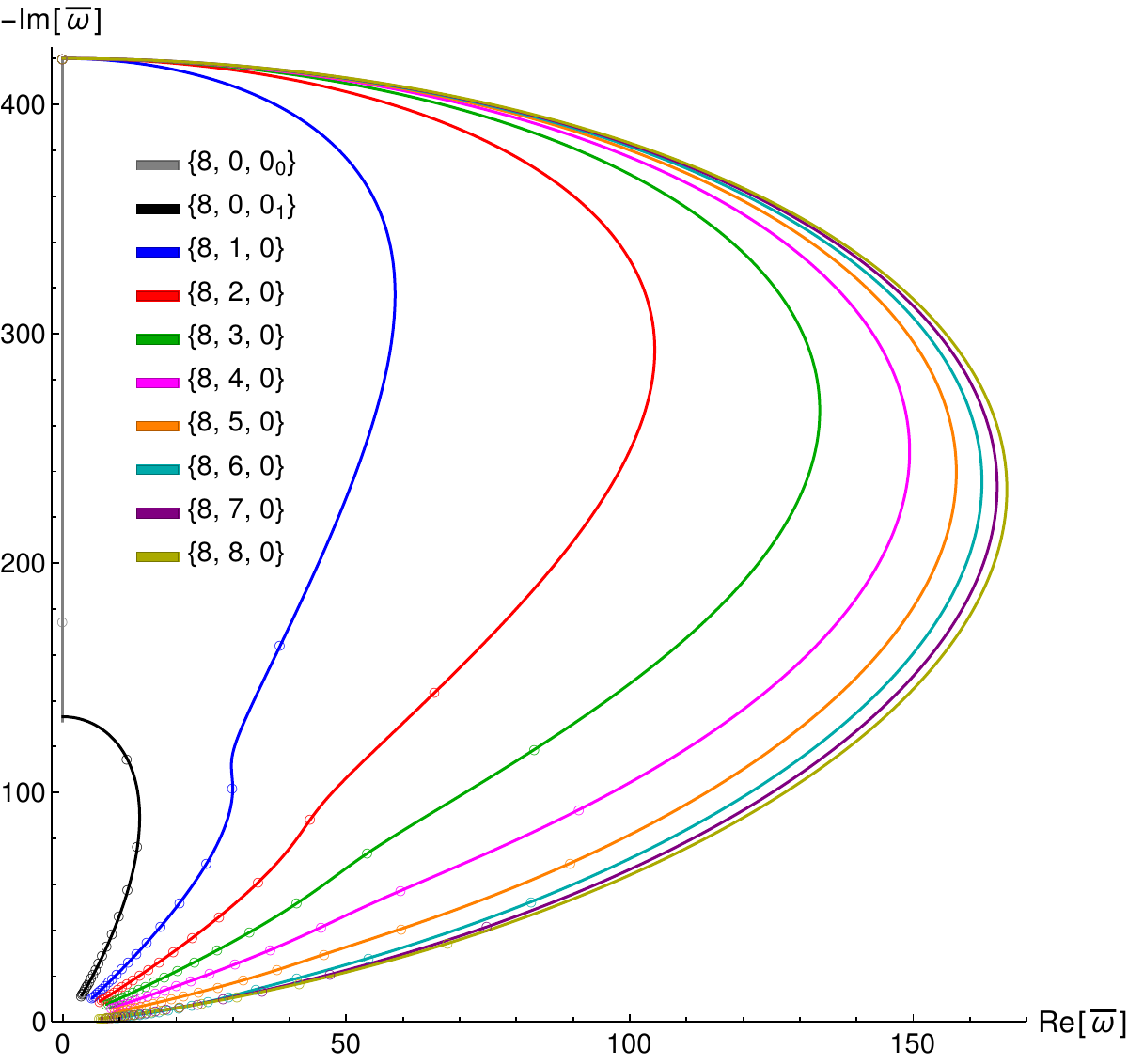} &
    \includegraphics[width=0.5\linewidth,clip]{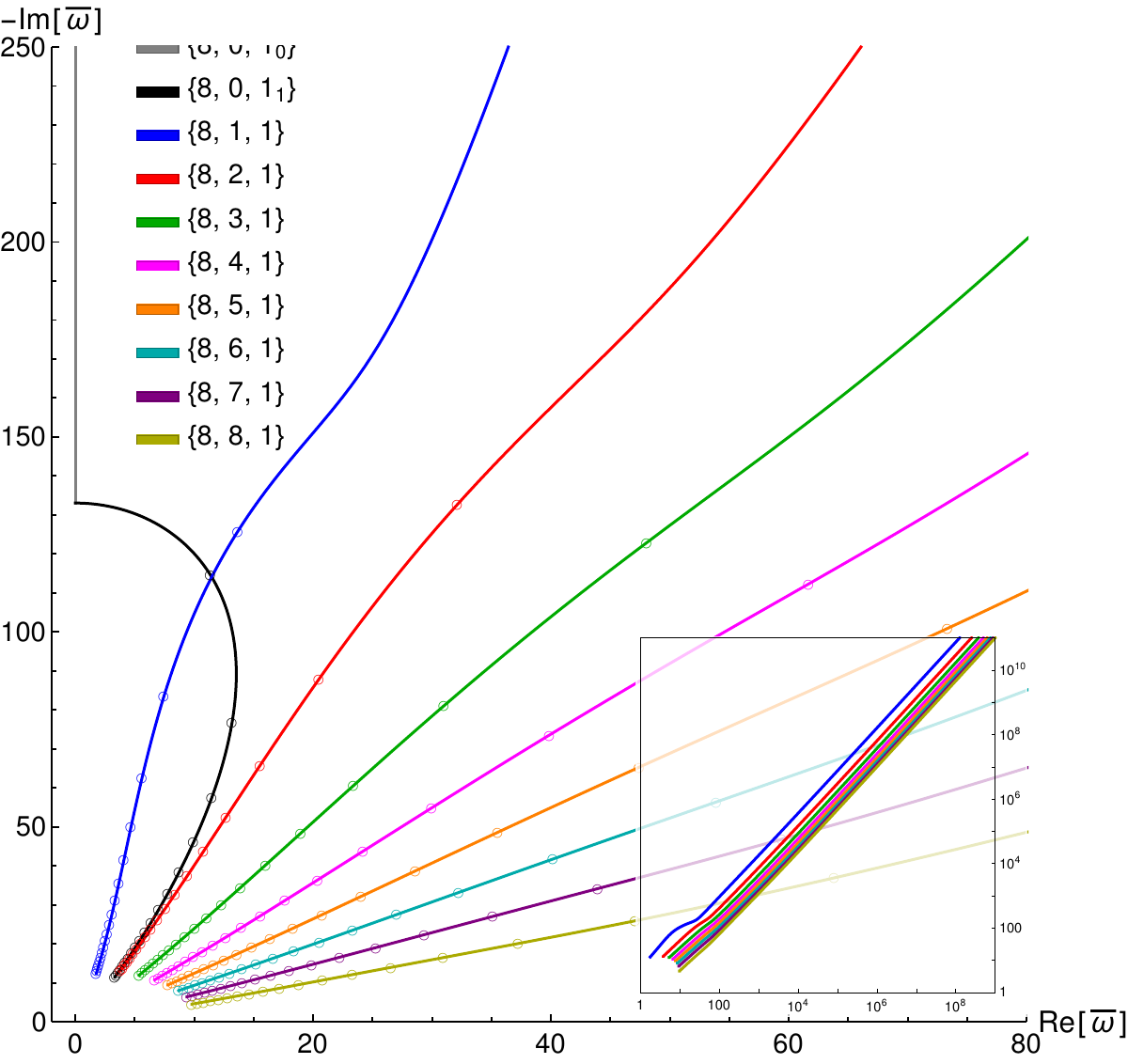} \\
    \multicolumn{2}{c}{\includegraphics[width=0.5\linewidth,clip]{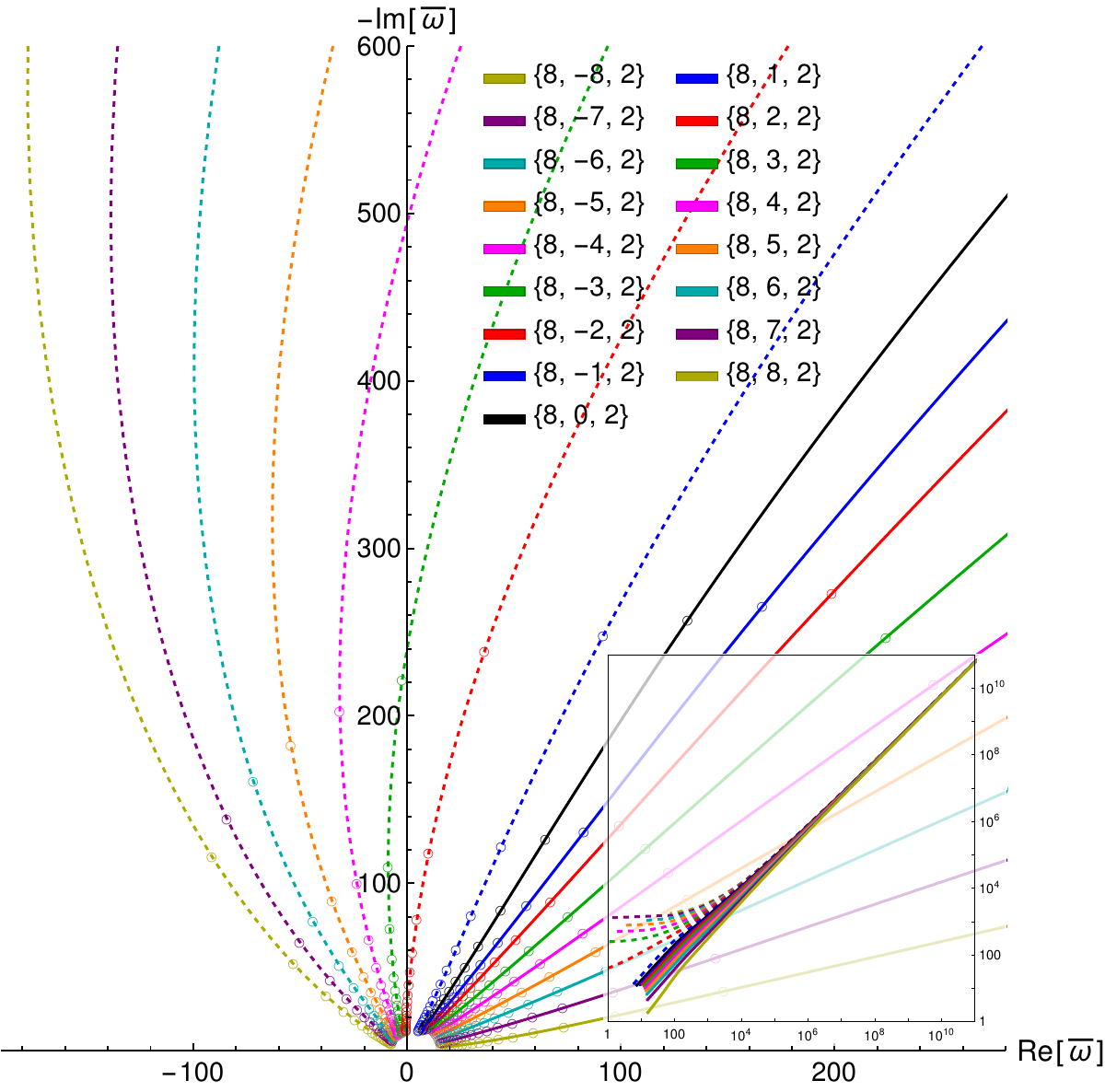}}
\end{tabular}
    \caption{Kerr TTM mode sequences for $\ell=8$. The original family denoted by $n=0$ is in the upper left plot.  The first new family denoted by $n=1$ is in the upper right plot.   The second new family denoted by $n=2$ is in the lower plot.  Mode sequences with negative values of $m$ are drawn as dashed lines.}
    \label{fig:l8 n012 all m}
\end{figure}

\end{widetext}


\begin{thebibliography}{24}%
\makeatletter
\providecommand \@ifxundefined [1]{%
 \@ifx{#1\undefined}
}%
\providecommand \@ifnum [1]{%
 \ifnum #1\expandafter \@firstoftwo
 \else \expandafter \@secondoftwo
 \fi
}%
\providecommand \@ifx [1]{%
 \ifx #1\expandafter \@firstoftwo
 \else \expandafter \@secondoftwo
 \fi
}%
\providecommand \natexlab [1]{#1}%
\providecommand \enquote  [1]{``#1''}%
\providecommand \bibnamefont  [1]{#1}%
\providecommand \bibfnamefont [1]{#1}%
\providecommand \citenamefont [1]{#1}%
\providecommand \href@noop [0]{\@secondoftwo}%
\providecommand \href [0]{\begingroup \@sanitize@url \@href}%
\providecommand \@href[1]{\@@startlink{#1}\@@href}%
\providecommand \@@href[1]{\endgroup#1\@@endlink}%
\providecommand \@sanitize@url [0]{\catcode `\\12\catcode `\$12\catcode
  `\&12\catcode `\#12\catcode `\^12\catcode `\_12\catcode `\%12\relax}%
\providecommand \@@startlink[1]{}%
\providecommand \@@endlink[0]{}%
\providecommand \url  [0]{\begingroup\@sanitize@url \@url }%
\providecommand \@url [1]{\endgroup\@href {#1}{\urlprefix }}%
\providecommand \urlprefix  [0]{URL }%
\providecommand \Eprint [0]{\href }%
\providecommand \doibase [0]{https://doi.org/}%
\providecommand \selectlanguage [0]{\@gobble}%
\providecommand \bibinfo  [0]{\@secondoftwo}%
\providecommand \bibfield  [0]{\@secondoftwo}%
\providecommand \translation [1]{[#1]}%
\providecommand \BibitemOpen [0]{}%
\providecommand \bibitemStop [0]{}%
\providecommand \bibitemNoStop [0]{.\EOS\space}%
\providecommand \EOS [0]{\spacefactor3000\relax}%
\providecommand \BibitemShut  [1]{\csname bibitem#1\endcsname}%
\let\auto@bib@innerbib\@empty
\bibitem [{\citenamefont {Kerr}(1963)}]{kerr-1963}%
  \BibitemOpen
  \bibfield  {author} {\bibinfo {author} {\bibfnamefont {R.~P.}\ \bibnamefont
  {Kerr}},\ }\bibfield  {title} {\bibinfo {title} {Gravitational field of a
  spinning mass as an example of algebraically special metrics},\ }\href
  {https://doi.org/10.1103/PhysRevLett.11.237} {\bibfield  {journal} {\bibinfo
  {journal} {Phys. Rev. Lett.}\ }\textbf {\bibinfo {volume} {11}},\ \bibinfo
  {pages} {237} (\bibinfo {year} {1963})}\BibitemShut {NoStop}%
\bibitem [{\citenamefont {Berti}\ \emph {et~al.}(2009)\citenamefont {Berti},
  \citenamefont {Cardoso},\ and\ \citenamefont {Starinets}}]{berti-QNM-2009}%
  \BibitemOpen
  \bibfield  {author} {\bibinfo {author} {\bibfnamefont {E.}~\bibnamefont
  {Berti}}, \bibinfo {author} {\bibfnamefont {V.}~\bibnamefont {Cardoso}},\
  and\ \bibinfo {author} {\bibfnamefont {A.~O.}\ \bibnamefont {Starinets}},\
  }\bibfield  {title} {\bibinfo {title} {Quasinormal modes of black holes and
  black branes},\ }\href {https://doi.org/10.1088/0264-9381/26/16/163001}
  {\bibfield  {journal} {\bibinfo  {journal} {Classical Quantum Gravity}\
  }\textbf {\bibinfo {volume} {26}},\ \bibinfo {pages} {163001} (\bibinfo
  {year} {2009})}\BibitemShut {NoStop}%
\bibitem [{\citenamefont {Konoplya}\ and\ \citenamefont
  {Zhidenko}(2011)}]{QNM-Review-2011}%
  \BibitemOpen
  \bibfield  {author} {\bibinfo {author} {\bibfnamefont {R.~A.}\ \bibnamefont
  {Konoplya}}\ and\ \bibinfo {author} {\bibfnamefont {A.}~\bibnamefont
  {Zhidenko}},\ }\bibfield  {title} {\bibinfo {title} {Quasinormal modes of
  black holes: From astrophysics to string theory},\ }\href
  {https://doi.org/10.1103/RevModPhys.83.793} {\bibfield  {journal} {\bibinfo
  {journal} {Rev. Mod. Phys.}\ }\textbf {\bibinfo {volume} {83}},\ \bibinfo
  {pages} {793} (\bibinfo {year} {2011})}\BibitemShut {NoStop}%
\bibitem [{\citenamefont {Abbott}\ \emph
  {et~al.}(2016{\natexlab{a}})\citenamefont {Abbott} \emph
  {et~al.}}]{GW150914-2016}%
  \BibitemOpen
  \bibfield  {author} {\bibinfo {author} {\bibfnamefont {B.~P.}\ \bibnamefont
  {Abbott}} \emph {et~al.} (\bibinfo {collaboration} {LIGO Scientific
  Collaboration and Virgo Collaboration}),\ }\bibfield  {title} {\bibinfo
  {title} {Observation of gravitational waves from a binary black hole
  merger},\ }\href {https://doi.org/10.1103/PhysRevLett.116.061102} {\bibfield
  {journal} {\bibinfo  {journal} {Phys. Rev. Lett.}\ }\textbf {\bibinfo
  {volume} {116}},\ \bibinfo {pages} {061102} (\bibinfo {year}
  {2016}{\natexlab{a}})}\BibitemShut {NoStop}%
\bibitem [{\citenamefont {Abbott}\ \emph
  {et~al.}(2016{\natexlab{b}})\citenamefont {Abbott} \emph
  {et~al.}}]{GW151226-2016}%
  \BibitemOpen
  \bibfield  {author} {\bibinfo {author} {\bibfnamefont {B.~P.}\ \bibnamefont
  {Abbott}} \emph {et~al.} (\bibinfo {collaboration} {LIGO Scientific
  Collaboration and Virgo Collaboration}),\ }\bibfield  {title} {\bibinfo
  {title} {{GW151226}: Observation of gravitational waves from a 22-solar-mass
  binary black hole coalescence},\ }\href
  {https://doi.org/10.1103/PhysRevLett.116.241103} {\bibfield  {journal}
  {\bibinfo  {journal} {Phys. Rev. Lett.}\ }\textbf {\bibinfo {volume} {116}},\
  \bibinfo {pages} {241103} (\bibinfo {year} {2016}{\natexlab{b}})}\BibitemShut
  {NoStop}%
\bibitem [{\citenamefont {Abbott}\ \emph
  {et~al.}(2017{\natexlab{a}})\citenamefont {Abbott} \emph
  {et~al.}}]{GW170104-2017}%
  \BibitemOpen
  \bibfield  {author} {\bibinfo {author} {\bibfnamefont {B.~P.}\ \bibnamefont
  {Abbott}} \emph {et~al.} (\bibinfo {collaboration} {LIGO Scientific
  Collaboration and Virgo Collaboration}),\ }\bibfield  {title} {\bibinfo
  {title} {{GW170104}: Observation of a 50-solar-mass binary black hole
  coalescence at redshift 0.2},\ }\href
  {https://doi.org/10.1103/PhysRevLett.118.221101} {\bibfield  {journal}
  {\bibinfo  {journal} {Phys. Rev. Lett.}\ }\textbf {\bibinfo {volume} {118}},\
  \bibinfo {pages} {221101} (\bibinfo {year} {2017}{\natexlab{a}})}\BibitemShut
  {NoStop}%
\bibitem [{\citenamefont {Abbott}\ \emph
  {et~al.}(2017{\natexlab{b}})\citenamefont {Abbott} \emph
  {et~al.}}]{GW170814-2017}%
  \BibitemOpen
  \bibfield  {author} {\bibinfo {author} {\bibfnamefont {B.~P.}\ \bibnamefont
  {Abbott}} \emph {et~al.} (\bibinfo {collaboration} {LIGO Scientific
  Collaboration and Virgo Collaboration}),\ }\bibfield  {title} {\bibinfo
  {title} {{GW170814}: A three-detector observation of gravitational waves from
  a binary black hole coalescence},\ }\href
  {https://doi.org/10.1103/PhysRevLett.119.141101} {\bibfield  {journal}
  {\bibinfo  {journal} {Phys. Rev. Lett.}\ }\textbf {\bibinfo {volume} {119}},\
  \bibinfo {pages} {141101} (\bibinfo {year} {2017}{\natexlab{b}})}\BibitemShut
  {NoStop}%
\bibitem [{\citenamefont {Wald}(1973)}]{wald-1973}%
  \BibitemOpen
  \bibfield  {author} {\bibinfo {author} {\bibfnamefont {R.~M.}\ \bibnamefont
  {Wald}},\ }\bibfield  {title} {\bibinfo {title} {On perturbations of a {K}err
  black hole},\ }\href {https://doi.org/10.1063/1.1666203} {\bibfield
  {journal} {\bibinfo  {journal} {J.~Math. Phys. (N.Y.)}\ }\textbf {\bibinfo
  {volume} {14}},\ \bibinfo {pages} {1453} (\bibinfo {year}
  {1973})}\BibitemShut {NoStop}%
\bibitem [{\citenamefont {Chandrasekhar}(1984)}]{chandra-1984}%
  \BibitemOpen
  \bibfield  {author} {\bibinfo {author} {\bibfnamefont {S.}~\bibnamefont
  {Chandrasekhar}},\ }\bibfield  {title} {\bibinfo {title} {On algebraically
  special perturbations of black holes},\ }\href
  {https://doi.org/10.1098/rspa.1984.0021} {\bibfield  {journal} {\bibinfo
  {journal} {Proc. R. Soc. A}\ }\textbf {\bibinfo {volume} {392}},\ \bibinfo
  {pages} {1} (\bibinfo {year} {1984})}\BibitemShut {NoStop}%
\bibitem [{\citenamefont {Cook}\ and\ \citenamefont
  {Zalutskiy}(2014)}]{cook-zalutskiy-2014}%
  \BibitemOpen
  \bibfield  {author} {\bibinfo {author} {\bibfnamefont {G.~B.}\ \bibnamefont
  {Cook}}\ and\ \bibinfo {author} {\bibfnamefont {M.}~\bibnamefont
  {Zalutskiy}},\ }\bibfield  {title} {\bibinfo {title} {Gravitational
  perturbations of the {K}err geometry: {H}igh-accuracy study},\ }\href
  {https://doi.org/10.1103/PhysRevD.90.124021} {\bibfield  {journal} {\bibinfo
  {journal} {Phys. Rev. D}\ }\textbf {\bibinfo {volume} {90}},\ \bibinfo
  {pages} {124021} (\bibinfo {year} {2014})}\BibitemShut {NoStop}%
\bibitem [{\citenamefont {Cook}\ \emph {et~al.}(2019)\citenamefont {Cook},
  \citenamefont {Annichiarico},\ and\ \citenamefont
  {Vickers}}]{cook-et-al-2018}%
  \BibitemOpen
  \bibfield  {author} {\bibinfo {author} {\bibfnamefont {G.~B.}\ \bibnamefont
  {Cook}}, \bibinfo {author} {\bibfnamefont {L.~S.}\ \bibnamefont
  {Annichiarico}},\ and\ \bibinfo {author} {\bibfnamefont {D.~J.}\ \bibnamefont
  {Vickers}},\ }\bibfield  {title} {\bibinfo {title} {Unknown branch of the
  total-transmission modes of the {K}err-geometry},\ }\href
  {https://doi.org/10.1103/PhysRevD.99.024008} {\bibfield  {journal} {\bibinfo
  {journal} {Phys. Rev. D}\ }\textbf {\bibinfo {volume} {99}},\ \bibinfo
  {pages} {024008} (\bibinfo {year} {2019})}\BibitemShut {NoStop}%
\bibitem [{\citenamefont {Andersson}(1994)}]{andersson-1994}%
  \BibitemOpen
  \bibfield  {author} {\bibinfo {author} {\bibfnamefont {N.}~\bibnamefont
  {Andersson}},\ }\bibfield  {title} {\bibinfo {title} {Total transmission
  through the {S}chwarzschild black-hole potential barrier},\ }\href
  {https://doi.org/10.1088/0264-9381/11/3/001} {\bibfield  {journal} {\bibinfo
  {journal} {Classical Quantum Gravity}\ }\textbf {\bibinfo {volume} {11}},\
  \bibinfo {pages} {L39} (\bibinfo {year} {1994})}\BibitemShut {NoStop}%
\bibitem [{\citenamefont {Keshet}\ and\ \citenamefont
  {Neitzke}(2008)}]{KeshetNeitzke2008}%
  \BibitemOpen
  \bibfield  {author} {\bibinfo {author} {\bibfnamefont {U.}~\bibnamefont
  {Keshet}}\ and\ \bibinfo {author} {\bibfnamefont {A.}~\bibnamefont
  {Neitzke}},\ }\bibfield  {title} {\bibinfo {title} {Asymptotic spectroscopy
  of rotating black holes},\ }\href
  {https://doi.org/10.1103/PhysRevD.78.044006} {\bibfield  {journal} {\bibinfo
  {journal} {Phys. Rev. D}\ }\textbf {\bibinfo {volume} {78}},\ \bibinfo
  {pages} {044006} (\bibinfo {year} {2008})}\BibitemShut {NoStop}%
\bibitem [{\citenamefont {Onozawa}(1997)}]{onozawa-1997}%
  \BibitemOpen
  \bibfield  {author} {\bibinfo {author} {\bibfnamefont {H.}~\bibnamefont
  {Onozawa}},\ }\bibfield  {title} {\bibinfo {title} {Detailed study of
  quasinormal frequencies of the {K}err black hole},\ }\href
  {https://doi.org/10.1103/PhysRevD.55.3593} {\bibfield  {journal} {\bibinfo
  {journal} {Phys. Rev. D}\ }\textbf {\bibinfo {volume} {55}},\ \bibinfo
  {pages} {3593} (\bibinfo {year} {1997})}\BibitemShut {NoStop}%
\bibitem [{\citenamefont {Cook}\ and\ \citenamefont
  {Zalutskiy}(2016)}]{cook-zalutskiy-2016b}%
  \BibitemOpen
  \bibfield  {author} {\bibinfo {author} {\bibfnamefont {G.~B.}\ \bibnamefont
  {Cook}}\ and\ \bibinfo {author} {\bibfnamefont {M.}~\bibnamefont
  {Zalutskiy}},\ }\bibfield  {title} {\bibinfo {title} {Modes of the {K}err
  geometry with purely imaginary frequencies},\ }\href
  {https://doi.org/10.1103/PhysRevD.94.104074} {\bibfield  {journal} {\bibinfo
  {journal} {Phys. Rev. D}\ }\textbf {\bibinfo {volume} {94}},\ \bibinfo
  {pages} {104074} (\bibinfo {year} {2016})}\BibitemShut {NoStop}%
\bibitem [{\citenamefont {Teukolsky}(1973)}]{teukolsky-1973}%
  \BibitemOpen
  \bibfield  {author} {\bibinfo {author} {\bibfnamefont {S.~A.}\ \bibnamefont
  {Teukolsky}},\ }\bibfield  {title} {\bibinfo {title} {Perturbations of a
  rotating black hole.\ {I}.\ {F}undamental equations for gravitational,
  electromagnetic, and neutrino-field perturbations},\ }\href
  {https://doi.org/10.1086/152444} {\bibfield  {journal} {\bibinfo  {journal}
  {Astrophys. J.}\ }\textbf {\bibinfo {volume} {185}},\ \bibinfo {pages} {635}
  (\bibinfo {year} {1973})}\BibitemShut {NoStop}%
\bibitem [{\citenamefont {Vickers}\ and\ \citenamefont
  {Cook}(2022)}]{VickersCook2022}%
  \BibitemOpen
  \bibfield  {author} {\bibinfo {author} {\bibfnamefont {D.~J.}\ \bibnamefont
  {Vickers}}\ and\ \bibinfo {author} {\bibfnamefont {G.~B.}\ \bibnamefont
  {Cook}},\ }\bibfield  {title} {\bibinfo {title} {Understanding solutions of
  the angular {T}eukolsky equation in the prolate asymptotic limit},\ }\href
  {https://doi.org/10.1103/PhysRevD.106.104037} {\bibfield  {journal} {\bibinfo
   {journal} {Phys. Rev. D}\ }\textbf {\bibinfo {volume} {106}},\ \bibinfo
  {pages} {104037} (\bibinfo {year} {2022})}\BibitemShut {NoStop}%
\bibitem [{\citenamefont {Ronveaux}(1995)}]{Heun-eqn}%
  \BibitemOpen
  \bibinfo {editor} {\bibfnamefont {A.}~\bibnamefont {Ronveaux}},\ ed.,\
  \href@noop {} {\emph {\bibinfo {title} {Heun's Differential Equations}}}\
  (\bibinfo  {publisher} {Oxford University, New York},\ \bibinfo {year}
  {1995})\BibitemShut {NoStop}%
\bibitem [{\citenamefont {Leaver}(1985)}]{leaver-1985}%
  \BibitemOpen
  \bibfield  {author} {\bibinfo {author} {\bibfnamefont {E.~W.}\ \bibnamefont
  {Leaver}},\ }\bibfield  {title} {\bibinfo {title} {An analytic representation
  for the quasi-normal modes of {K}err black holes},\ }\href
  {https://doi.org/10.1098/rspa.1985.0119} {\bibfield  {journal} {\bibinfo
  {journal} {Proc. R. Soc. A}\ }\textbf {\bibinfo {volume} {402}},\ \bibinfo
  {pages} {285} (\bibinfo {year} {1985})}\BibitemShut {NoStop}%
\bibitem [{\citenamefont {Casals}\ and\ \citenamefont
  {Ottewill}(2005)}]{Casalas-oblate-2005}%
  \BibitemOpen
  \bibfield  {author} {\bibinfo {author} {\bibfnamefont {M.}~\bibnamefont
  {Casals}}\ and\ \bibinfo {author} {\bibfnamefont {A.~C.}\ \bibnamefont
  {Ottewill}},\ }\bibfield  {title} {\bibinfo {title} {High frequency
  asymptotics for the spin-weighted spheroidal equation},\ }\href
  {https://doi.org/10.1103/PhysRevD.71.064025} {\bibfield  {journal} {\bibinfo
  {journal} {Phys. Rev. D}\ }\textbf {\bibinfo {volume} {71}},\ \bibinfo
  {pages} {064025} (\bibinfo {year} {2005})}\BibitemShut {NoStop}%
\bibitem [{\citenamefont {Berti}\ \emph {et~al.}(2006)\citenamefont {Berti},
  \citenamefont {Cardoso},\ and\ \citenamefont
  {Casals}}]{berticardosocasals-2006}%
  \BibitemOpen
  \bibfield  {author} {\bibinfo {author} {\bibfnamefont {E.}~\bibnamefont
  {Berti}}, \bibinfo {author} {\bibfnamefont {V.}~\bibnamefont {Cardoso}},\
  and\ \bibinfo {author} {\bibfnamefont {M.}~\bibnamefont {Casals}},\
  }\bibfield  {title} {\bibinfo {title} {Eigenvalues and eigenfunctions of
  spin-weighted spheroidal harmonics in four and higher dimensions},\ }\href
  {https://doi.org/10.1103/PhysRevD.73.024013} {\bibfield  {journal} {\bibinfo
  {journal} {Phys. Rev. D}\ }\textbf {\bibinfo {volume} {73}},\ \bibinfo
  {pages} {024013} (\bibinfo {year} {2006})}\BibitemShut {NoStop}%
\bibitem [{\citenamefont {Barrowes}\ \emph {et~al.}(2004)\citenamefont
  {Barrowes}, \citenamefont {O'Neill}, \citenamefont {Grzegorczyk},\ and\
  \citenamefont {Kong}}]{Barrowes-etal-2004}%
  \BibitemOpen
  \bibfield  {author} {\bibinfo {author} {\bibfnamefont {B.~E.}\ \bibnamefont
  {Barrowes}}, \bibinfo {author} {\bibfnamefont {K.}~\bibnamefont {O'Neill}},
  \bibinfo {author} {\bibfnamefont {T.~M.}\ \bibnamefont {Grzegorczyk}},\ and\
  \bibinfo {author} {\bibfnamefont {J.~A.}\ \bibnamefont {Kong}},\ }\bibfield
  {title} {\bibinfo {title} {On the asymptotic expansion of the spheroidal wave
  function and its eigenvalues for complex size parameter},\ }\href
  {https://doi.org/https://doi.org/10.1111/j.0022-2526.2004.01526.x} {\bibfield
   {journal} {\bibinfo  {journal} {Studies in Applied Mathematics}\ }\textbf
  {\bibinfo {volume} {113}},\ \bibinfo {pages} {271} (\bibinfo {year}
  {2004})}\BibitemShut {NoStop}%
\bibitem [{\citenamefont {Hod}(1998)}]{Hod-1998}%
  \BibitemOpen
  \bibfield  {author} {\bibinfo {author} {\bibfnamefont {S.}~\bibnamefont
  {Hod}},\ }\bibfield  {title} {\bibinfo {title} {Bohr's correspondence
  principle and the area spectrum of quantum black holes},\ }\href
  {https://doi.org/10.1103/PhysRevLett.81.4293} {\bibfield  {journal} {\bibinfo
   {journal} {Phys. Rev. Lett.}\ }\textbf {\bibinfo {volume} {81}},\ \bibinfo
  {pages} {4293} (\bibinfo {year} {1998})}\BibitemShut {NoStop}%
\bibitem [{DEA(2021)}]{DEAC-Cluster}%
  \BibitemOpen
  \href {https://doi.org/10.57682/G13Z-2362} {\bibinfo {title} {{WFU High
  Performance Computing Facility}}} (\bibinfo {year} {2021})\BibitemShut
  {NoStop}%
\end{thebibliography}
\end{document}